\documentclass[twocolumn,prc,floatfix,preprintnumbers,nofootinbib,superscriptaddress]{revtex4-2}

\usepackage{color} 

\usepackage{graphicx}

\usepackage{isotope}
\usepackage{physics}

\def\SCNTHuizhou{Southern Center for Nuclear-Science Theory (SCNT), Institute of Modern Physics, Chinese Academy of Sciences, Huizhou 516000, China}
\def\wigner{Wigner Research Centre for Physics, Budapest, 1121, Hungary}
\def\kyiv{Institute for Nuclear Research, National Academy of Sciences of Ukraine, Kyiv, 03680, Ukraine}
\def\HISTKLLanzhou{Heavy Ion Science and Technology Key Laboratory, Institute of Modern Physics, Chinese Academy of Sciences, Lanzhou 730000, China}
\def\UnivCASBeijing{School of Nuclear Sciences and Technology, University of Chinese Academy of Sciences, Beijing 101408, China}

\begin{document}

\title{Bremsstrahlung emission accompanying ternary fission of \isotope[252]{Cf}}

\author{Sergei~P.~Maydanyuk}~\email{sergei.maydanyuk@impcas.ac.cn} 
\affiliation{\SCNTHuizhou}
\affiliation{\wigner}
\affiliation{\kyiv}

\author{Ju-Jun Xie}~\email{xiejujun@impcas.ac.cn}
\affiliation{\SCNTHuizhou}
\affiliation{\HISTKLLanzhou}
\affiliation{\UnivCASBeijing}

\author{Sergei~O.~Omelchenko}~\email{sergomel@ukr.net}
\affiliation{\kyiv}


\date{\small\today}

\begin{abstract}
%
\begin{description}
\item[Background]
In ternary fission, bremsstrahlung photons are emitted but those have never been studied yet theoretically and experimentally.
In other reactions, bremsstrahlung has been studied for a long time.


\item[Purpose]
To clarify which new information about ternary fission can be obtained from study of
bremsstrahlung emission accompanying the ternary fission of \isotope[252]{Cf}.


\item[Methods]
A new quantum model of emission of bremsstrahlung photons accompanying ternary fission of heavy nuclei
with $\alpha$-particle as light charged particle is developed. 
The model takes into account geometry and dynamics of ternary fission.
%
\item[Results]
We present the theoretical results on the bremsstrahlung emission in the ternary fission of the \isotope[252]{Cf} nucleus. High sensitivity of the bremsstrahlung spectra is established by the model concerning to the following aspects of the ternary fission, and the theoretical calcualtions are in agreement with the preliminary experimental data. It is found that:
(a) Photons are emitted with highest intensity in case of perpendicular motion of the $\alpha$\,particle concerning to fission axis;
(b) Relative motion between heavy fragments reinforces significantly bremsstrahlung, leaving of $\alpha$-particle concerning to system of heavy fragments is less important;
(c) Relative motion between two heavy fragments is faster, bremsstrahlung is more intensive.

%


\item[Conclusions]
Theoretical study of bremsstrahlung in ternary fission of the \isotope[252]{Cf} nucleus shows high sensitivity of bremsstrahlung spectra on the geometry and dynamics of ternary fission process. It is expected that new information can be obtained by the model when new experimental measurements of bremsstrahlung in ternary fission of the \isotope[252]{Cf} nucleus are available.
\end{description}


\end{abstract}




\maketitle


\section{Introduction
\label{sec.Introduction}}


Spontaneous fission accompanied by additional emission of the light charged particle (LCP) is called as \emph{ternary fission}, which is important for nuclear dynamics. Ternary fission is a rare decay process (less than 1 percent, $\simeq 1/260$ for \isotope[252]{Cf}) comparing with to binary fission~\cite{Kopatch:2002bd}. About 87\% of ternary fission events with emission of $\alpha$-particle is present~\cite{Kopatch:2002bd}. And the process is characterized by a continuous (near-Gaussian) energy spectrum with $\simeq 16$~MeV mean energy and $\simeq 11$~MeV width. As is known, the probability of such a process decreases sharply with increasing mass number of LCP (see yields of ternary particles in the ternary fission relative to binary fission for \isotope[236]{U}, \isotope[243]{Am}, \isotope[250]{Cf} in Fig.~1 of Ref.~\cite{Zagrebaev:2010jv}, LCPs from neutron till nucleus with mass number $A=38$ are used in analysis in this paper, see also Ref.~\cite{Gonnenwein.1999.chapter}). Theoretical explanation gives interpretation of such a phenomenon as an indication that LCP is emitted from the neck region between two heavy fragments. Essential forces were given to the experimental study of the ternary fission~\cite{Schall:1987hwr,Daniel:2004jx,Ramayya07,Ramayya:1998zza,kopach,Mutterer:2008zz,Vermote:2010gnf,Pyatkov:2017afj}.
In Ref.~\cite{kopach} LCPs \isotope[3]{H}, \isotope[4]{He}, \isotope[6]{He}, Li, and Be accompanied fission were studied experimentally. In that work, the angular distributions of $\gamma$-ray were also deduced with respect to the motion of both of the fragments and LCPs. Heavier clusters like \isotope[10]{Be}, \isotope[14]{C}, \isotope[20]{O}, \isotope[24]{Ne}, \isotope[28]{Mg}, and \isotope[34]{Si} have also been detected~\cite{Gonnenwein.1993.ConfSer}. Indeed, LCP with the highest probability is $\alpha$\,particle~\cite{Ramayya:1998zza,kopach}. Other LCPs with still sizable partial yields are \isotope[3]{H} ($\simeq 7$~\%), the neutron-rich \isotope[]{He} isotopes \isotope[6]{He} ($\simeq 3.5$~\%) and
\isotope[8]{He} ($\simeq 0.2$~\%), the \isotope[7,8,9]{Li} ($\simeq 0.5$~\%) and \isotope[9,10,11]{Be} ($\simeq 2$~\%) nuclei~\cite{Kopatch:2002bd}.

On the theoretical side, ternary fission has been investigated on the basis of liquid drop models (LDM)~\cite{Swiatecki.1958.conf.p651,Diehl:1974uol,Royer:2024mag}, three-center shell models~\cite{Degheidy.1979.ZPhysA,Karpov:2016ucy}, three-body shell models~\cite{Denisov:2022cqn,Denisov:2017ykd}, two-center shell model~\cite{Zagrebaev:2010jv}, improved scission point model~\cite{Ivanyuk:2024ssm}, quantum mechanical models~\cite{Oberstedt_Carjan.1992.ZPhysA},  non-equilibrium information entropy approach~\cite{Ropke:2020hbm,Ropke:2020fut}, approaches of three-cluster model potential energy surfaces~\cite{Balasubramaniam:2016oby}, semiempirical approaches on the basis of $\alpha$-decay properties of fissioning nuclei~\cite{Khuyagbaatar:2024sun}, and Monte Carlo studies of $\alpha$-accompanied fission~\cite{Radi:1982zz}.

In addition to the process above, there is another type of ternary fission process 
where heavy nucleus is separated simultaneously on three fragments of similar not light masses.
Such a process is obtained term as \emph{``true ternary fission''}%
~\cite{Zagrebaev:2010jv,Tashkhodjaev:2023jhq,%
vonOertzen:2020jem}.
History of investigations of true ternary fission is long in theoretical and even experimental study
(even in 50-th of past century it was known about true ternary fission \cite{PhysRev.78.533}). From LDM for the ternary fission
it becomes clear that barriers for oblate (triangle) deformations are much larger
than the barriers of prolate configurations~\cite{Diehl:1974uol}. Moreover, such barriers with oblate (triangle) deformations for true ternary fission are much larger than for the binary fission. In this approach it was found that ternary fission produces larger total energy release in comparison with to the binary fission~\cite{Swiatecki.1958.conf.p651}. 
From early study it was concluded that main aspects in true ternary fission are properties of fission barriers and not total energy release.
However, three-center shell model~\cite{Degheidy.1979.ZPhysA} 
showed that the shell effects significantly reduce barriers for the ternary fission.



In the framework of LDM, it was found for those nuclei that 
shells corrections to total deformation energy are much more important than fission barriers which are enough low.
However, further theoretical study found nuclei, for example \isotope[298]{114}, 
where fission barriers (with addition of shell corrections) 
play important role \cite{Schultheis:1974kra}. 
This also reinforces interest on quantum mechanical study of these processes. Moreover, decay onto three doubly magic heavy fragments might occur for giant nuclear systems 
which can be formed in the low-energy collisions of actinide heavy nuclei~\cite{Zagrebaev:2010jv}. 
It was found that compound nucleus is hardly be formed, and such a decay is quasi-fission process.
Stability of superheavy nuclei with different arrangements of the fragments in ternary fission has been studied by
approaches of model potential energy surfaces~\cite{Balasubramaniam:2016oby}.

Ternary fission is accompanied by the addition emission of neutrons and bremsstrahlung photons. 
Such bremsstrahlung photons can be measured experimentally and provide us additional more rich information about ternary fission.
As in Ref.~\cite{Maydanyuk_2011}, this emission in the ternary fission has never been studied yet, both theoretically and experimentally.
However, investigations of such photons in nuclear reactions is traditional topic of nuclear physics existed for a long time.
Some preliminary results of a theoretical study of bremsstrahlung in the ternary fission of \isotope[252]{Cf} were presented in Ref.~\cite{Maydanyuk_2011}. Those results are needed in more systematic and deep analysis. Hence, we will investigate the ternary fission of $^{252}$Cf with $\alpha$-particle as LCP in this work.

The paper is organized in the following way.
In Sec.~\ref{sec.2} a new model of emission of the bremsstrahlung photons in the ternary fission with $\alpha$-particle as LCP is presented. 
Here, geometry of fissioning nuclear system is fixed, 
bremsstrahlung matrix elements and corresponding probabilities for different processes in ternary fission are formulated, 
full bremsstrahlung probability is defined.
In Sec.~\ref{sec.3} the bremsstrahlung probabilities for different processes in the ternary fission of \isotope[252]{Cf} are calculated with the corresponding analysis. Bremsstrahlung spectrum in $\alpha$ decay of \isotope[252]{Cf} is calculated.
To test model, this result is compared with well established results in study of bremsstrahlung in $\alpha$ decay of 
\isotope[214]{Po}, \isotope[226]{Ra}~\cite{Giardina:2008pwx,Giardina:2008sd}. In the ternary fission, spectra are calculated concerning to different mass separations of two heavy fragments, different angular geometries of leaving between heavy fragments and $\alpha$ particle. Different relative motions (i.e., dynamics) between heavy fragments and $\alpha$\,particle are studied with calculations of the corresponding spectra. Role of neck between two heavy fragments at moment of separation of $\alpha$ particle is discussed. Finally, the full spectrum is calculated in comparison with preliminary experimental data~\cite{Maydanyuk_2011}. Conclusions are summarized in Sec.~\ref{sec.conclusions}.
Useful details of calculation of the bremsstrahlung matrix element from contribution caused by the mass separation of the binary nucleus
are added in Appendix~\ref{sec.app.1}.


\section{Model of bremsstrahlung emission accompanying ternary fission
\label{sec.2}}

A specific issue of the ternary fission is that three nuclear fragments participate in this process, 
and relative distances between those are changed during fission.
In the case of two fragments (i.e., for processes like $\alpha$-decay or spontaneous fission of nucleus on two heavy fragments) 
the matrix element of emission is calculated on the basis of the stationary wave functions of the decaying or fissioning nuclear system. 
However, if at leaving of $\alpha$-particle the binary nucleus is additionally separated on two heavy fragments,  
then the interaction potential between those and $\alpha$-particle is dependent on 
geometry of separation between each others and relative velocities of motion of all fragments. 
Therefore, the matrix element of bremsstrahlung should be defined taking into account all these aspects of geometry and dynamics of the ternary fission.

\subsection{Geometry of the surface of the  nuclear system composed of three fragments
\label{sec.2.1.1}}

To develop a new model describing bremsstrahlung in the ternary fission, it is convenient to use the bremsstrahlung model for the spontaneous fission in the basis. In Ref.~\cite{Maydanyuk:2010zz}, the bremsstrahlung emission in the spontaneous fission of the \isotope[252]{Cf} nucleus was investigated with quantum mechanical model in the basis. The theoretical results are in agreement with the available experimental data. Here, we will improve the model with inclusion of formalism for the bremsstrahlung in the ternary fission with the following steps:
%
(a) To define the geometry of the surface of the nuclear system undergoing the ternary fission.
%
(b) To calculate the interaction potential between LCP and
binary nucleus composed of two heavy fragments 
taking different geometrical arrangements of fragments and LCP into account.
(c) To calculate the matrix element of bremsstrahlung taking motion of two fragments of the binary nucleus into account.
%
(d) 
Calculations of the spectra should be stable 
for different separations of the binary nucleus on the light and heavy fragments, for different geometries of the fission.

\begin{figure*}[htbp]
\centering
\vspace{-5cm}
\includegraphics[scale=0.4]{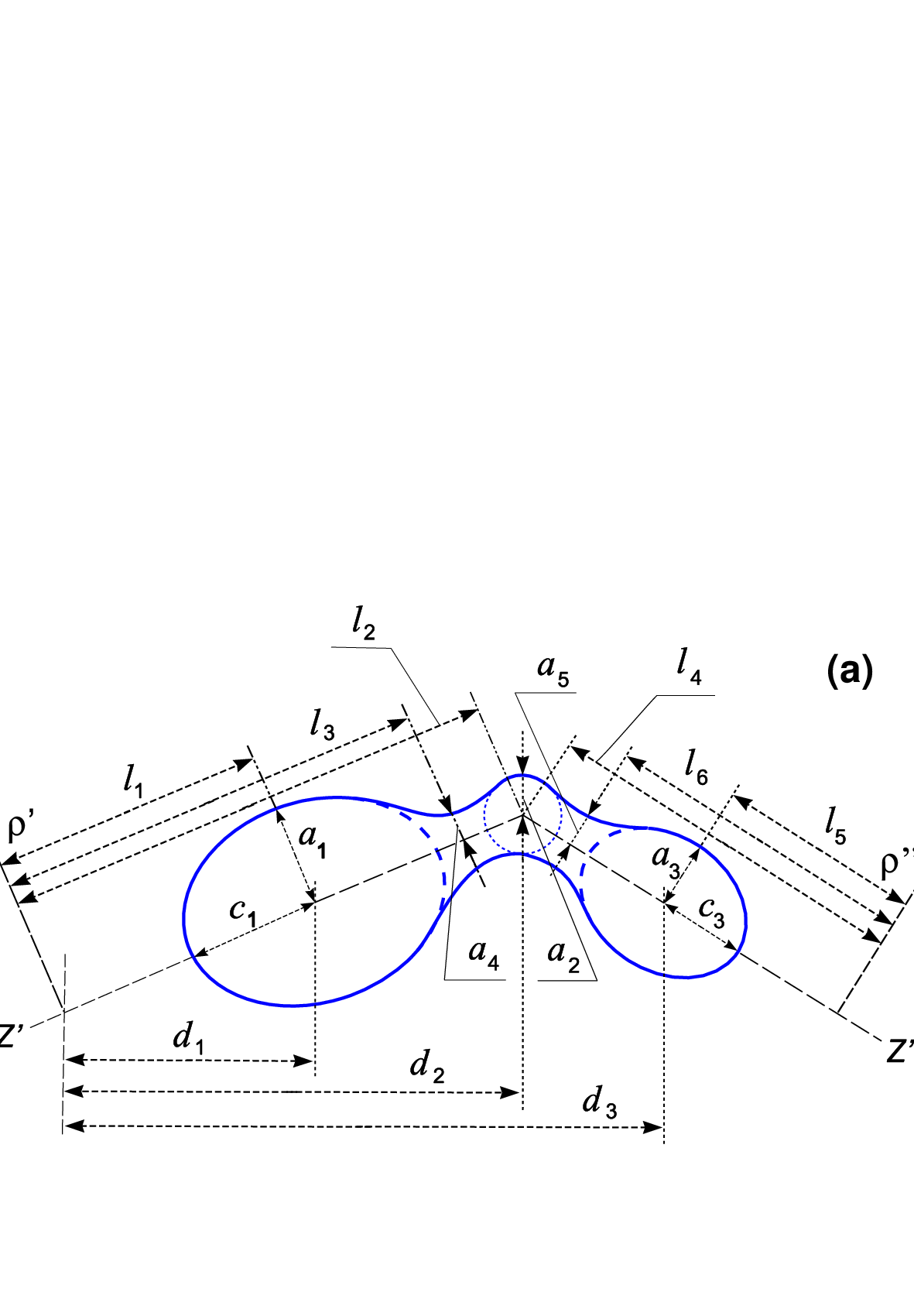}
\includegraphics[scale=0.4]{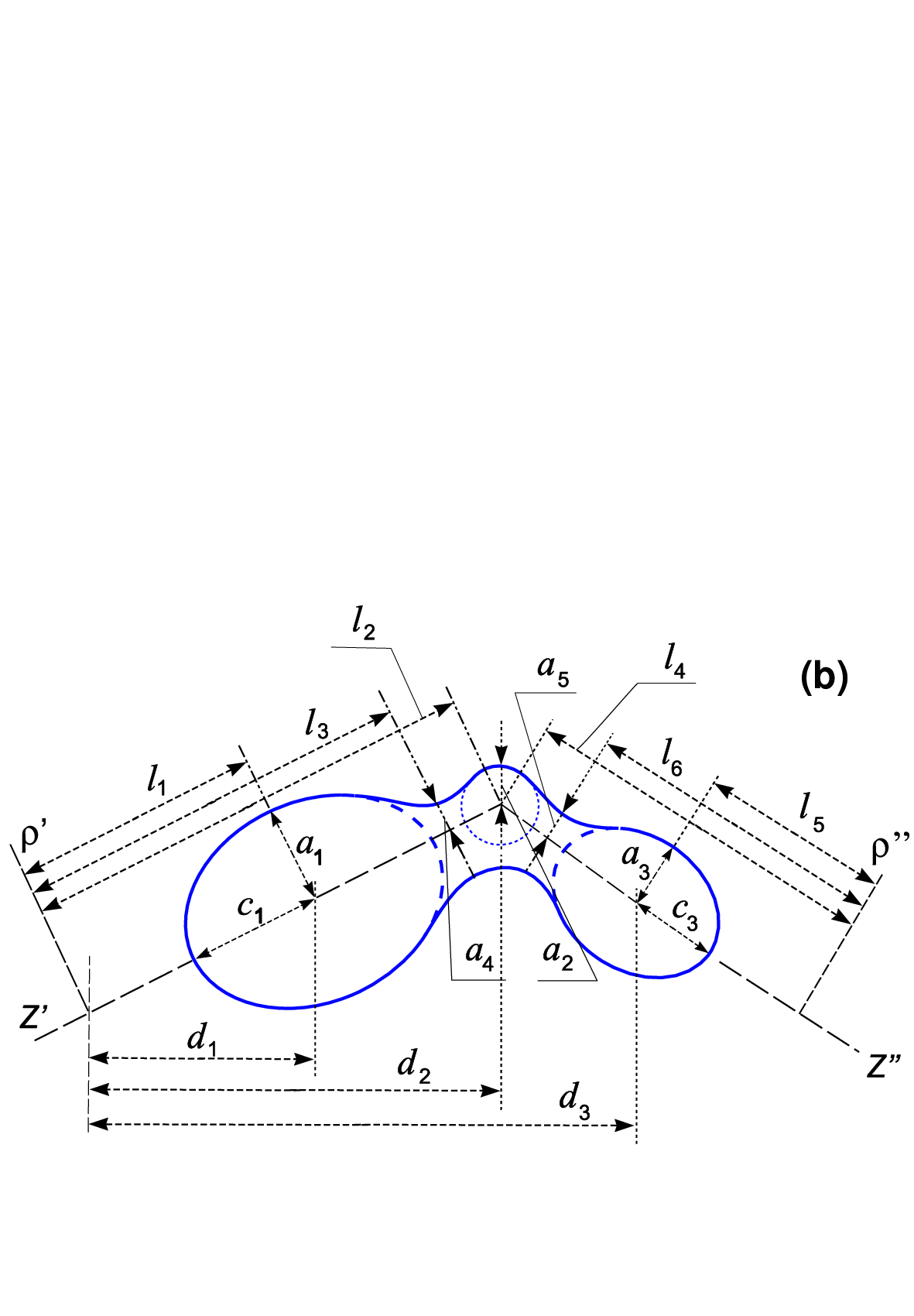}
\vspace{-1cm}
\caption{\small (Color online) Scheme of the nuclear system undergoing the ternary fission:
(a) at far distance between two fragments of the binary nucleus the $\alpha$\,particle is connected with each fragment through corresponding neck;
(b) for a small distance between two heavy fragments of the binary nucleus, at geometrical region opposite to the emission of the $\alpha$-particle, the surfaces of the heavy fragments are connected by the joint hyperbolic surface.
\label{fig.1}}
\end{figure*}

Following Ref.~\cite{Bolsterli:1972zz}, we have defined the nuclear shapes and calculated potential for separation of nucleus \isotope[252]{Cf} on two fragments in the spontaneous fission~~\cite{Maydanyuk_2011,Maydanyuk:2010zz}. To apply the approach for the ternary fission, we further define the geometry of the surface of the  nuclear system composed of three fragments as (see more details in Fig.~\ref{fig.1})
\begin{equation}
\begin{array}{lll}
\vspace{1.5mm}
  \mbox{\rm for } z \le d_{2}: \\

\vspace{2.5mm}  (\rho^{\prime})^{2} =
\left\{
\begin{array}{ll}
  \vspace{1mm}
  a_{1}^{2} - (a_{1}^{2}/c_{1}^{2})\: (z^{\prime}-l_{1})^{2} &
  \mbox{at } l_{1}-c_{1} \le z^{\prime} \le z_{1}, \\
  \vspace{1mm}
  a_{2}^{2} -
  (z^{\prime}-l_{2})^{2} &
  \mbox{at } z_{2} \le z^{\prime} \le l_{2}+c_{2}, \\
  a_{4}^{2} + (a_{4}^{2}/c_{4}^{2})\: (z^{\prime}-l_{3})^{2} &
  \mbox{at } z_{1} \le z^{\prime} \le z_{2},
\end{array} \right. \\

\vspace{1.5mm}
\mbox{\rm for } z \ge d_{2}: \\
  (\rho^{\prime\prime})^{2} =
\left\{
\begin{array}{ll}
  \vspace{1mm}
  a_{3}^{2} -
  (a_{3}^{2}/c_{3}^{2})\: (z^{\prime\prime}-l_{5})^{2} &
  \mbox{at } l_{5}-c_{3} \le z^{\prime\prime} \le z_{3}, \\
  \vspace{1mm}
  a_{2}^{2} -
  (z^{\prime\prime}-l_{4})^{2} &
  \mbox{at } z_{4} \le z^{\prime\prime} \le l_{4}+c_{2}, \\
  a_{5}^{2} +
  (a_{5}^{2}/c_{5}^{2})\: (z^{\prime\prime}-l_{6})^{2} &
  \mbox{at } z_{3} \le z^{\prime\prime} \le z_{4}.
\end{array} \right. \\
\end{array}
\label{eq.2.1.2.1}
\end{equation}
The approximation of a spherically symmetric LCP ($a_{2}=c_{2}$) is used in this set of relations.
Fig.~\ref{fig.1}~(a) shows the scheme of the parameters for the nuclear system composed of three fragments which are connected by two necks.
As is known, LCP starts to separate (and move outside) at a small enough distance between two heavy fragments.
So, at the geometrical region opposite to separation of LCP, surfaces of the heavy fragments are connected by a joint hyperbolic surface which does not touch LCP [see Fig.~\ref{fig.1}~(b)].
The necks are formed at begining of the relative motion of fragments.

The helium as LCP is present in over 90~\% of all ternary fission events~~\cite{Hwang:2000nt,Mutterer.1996.JINR}, we shall choose $\alpha$\,particle as LCP in this work. The $\alpha$\,particle is emitted with the highest probability nearly perpendicularly with respect to the fission axis. It is usually suggested to be emitted from the neck region. We take this into account in the description of the surface geometry of the ternary fissioning nuclear system (see Fig.~\ref{fig.2}) and in calculation of the potential. In such a basis, we shall assume that the $\alpha$\,particle begins to separate from location on the fission axis.


\subsection{The potential of interaction between the $\alpha$-particle and the binary nucleus
\label{sec.2.2}}

Once the nuclear surface has been specified, the next step is to generate an interaction potential between the binary nucleus and the $\alpha$-particle. Such a potential is written down as
\begin{equation}
\begin{array}{ccl}
  V_{\rm total}(\mathbf{r}) & = &
  V_{\rm N}(\mathbf{r}) + V_{\rm C}(\mathbf{r}),
\end{array}
\label{eq.2.2.1.1}
\end{equation}
where $\mathbf{r}$ is radius-vector from center-of-mass of the binary nucleus to center-of-mass of the $\alpha$\,particle. 
The nuclear component $V_{\rm N}(\mathbf{r})$ and
the Coulomb component $V_{\rm C}(\mathbf{r})$ 
for the binary fission are defined 
as~\cite{Maydanyuk:2010zz}
\begin{equation}
\begin{array}{cclccl}
\vspace{1.5mm}
V_{\rm N}(\mathbf{r}) & = & E_{\rm N,\, nucleus}(\mathbf{r}) - E_{\rm N,\, fragment}\,, \\

  V_{\rm C}(\mathbf{r}) & = & E_{\rm C,\, nucleus}(\mathbf{r}) - E_{\rm C,\, fragment}\,,
\end{array}
\label{eq.2.2.1.2}
\end{equation}
where $E_{\rm N,\, nucleus}(\mathbf{r})$ and $E_{\rm C,\, nucleus}(\mathbf{r})$ are nuclear and Coulomb potential energies, respectively, for the full nuclear system, 
while $E_{\rm N,\, fragment}(\mathbf{r})$ and $E_{\rm C,\, fragment}(\mathbf{r})$ are nuclear and Coulomb potential energies, respectively, 
for the emitted 
fragment~\cite{Maydanyuk:2010zz,Bolsterli:1972zz}:
\begin{equation}
\begin{array}{lll}
  E_{\rm N,\, nucleus}(\mathbf{r}) =
    -\lambda_{N}\;
    \displaystyle\int\limits_{V}
      \displaystyle\frac{\mathbf{dr^{\prime}}^{3}}{1 + \exp(|\mathbf{r} - \mathbf{r}^{\prime}|/a)}\,,  \\
      
  E_{\rm N,\, fragment} =
    -\lambda_{N}\;
    \displaystyle\int\limits_{V}
      \displaystyle\frac{\mathbf{dr}^{3}}{1 + \exp(|\mathbf{r}|/a)}\,, \\
  E_{\rm C,\, nucleus}(\mathbf{r}) =
    \lambda_{C}\, \displaystyle\int\limits_{V,\, \mathbf{r} \ne \mathbf{r^{\prime}}}
      \displaystyle\frac{\mathbf{dr^{\prime}}^{3}}{|\mathbf{r}^{\prime} - \mathbf{r}|}\,,  \\

  E_{\rm C,\, fragment} =
    \lambda_{C}\, \displaystyle\int\limits_{V,\, r \ne 0}
      \displaystyle\frac{\mathbf{dr}^{3}}{|\mathbf{r}|}\,.
\end{array}
\label{eq.2.2.1.3}
\end{equation}
The integration is over the volume $V$ of the fissioning nuclear system bounded by the given surface (defined concerning to the distance
$r$ between centers-of-masses of the binary nucleus and $\alpha$ particle). 
Parameters $\lambda_{C} = Z_{1} Z_{2}\,e^{2} / V_{p}$ and $\lambda_{N} = M_{p}/V_{p}$ 
(where $V_{p}$ and $M_{p}$ are volume and mass of the parent nucleus, respectively) give similar results for the potential 
calculated by Eqs.~(\ref{eq.2.2.1.1})--(\ref{eq.2.2.1.3}) and the potential calculated
by the procedure with parametrization of Denisov and Ikezoe in Ref.~\cite{Denisov:2005ax} for $\alpha$-decay of \isotope[252]{Cf}.


The approach to calculate the potential allows to study much more complicated deformations of the nuclear system than binary separation of the nucleus in the spontaneous fission. 
Thus, in order to study the ternary fission, we should describe separation of the binary nucleus on two fragments with different masses. Using the above-defined surface of such a nuclear system composed of three fragments, we 
determine a new potential $V_{\rm total}$ between the $\alpha$\,particle and   binary nucleus when it is still not separated into two fragments. In general, such a potential is already dependent on the much more
complicated geometry of the relative location of deformed fragments with different sizes and orientations.

In order to realize practically the first calculation of  bremsstrahlung spectra accompanying such a ternary process,
we shall use the spherically symmetric approximation of the two separated fragments of the  binary nucleus [$a_{1}=c_{1}$ and $a_{3}=c_{3}$ in Figs. \ref{fig.1} (a) and (b)]. As main parameters of the defined geometry of fission, we shall choose
(i) the relative distance $r$ between the centers of masses of the $\alpha$\,particle and  binary nucleus,
(ii) the relative distance $R_{12}$ between the centers of masses of the two fragments of the binary nucleus, and
(iii) the $\varphi$ angle between the line connecting these two fragments and the $\mathbf{r}$ vector directed from the center of mass of the binary nucleus to the center of mass of the $\alpha$\,particle. Then, the total potential is written down as
\begin{equation}
\begin{array}{ccl}
  V_{\rm total}(\mathbf{r}, R_{12}, \varphi) & = &
  V_{\rm N}(\mathbf{r}, R_{12}, \varphi) + V_{\rm C}(\mathbf{r}, R_{12}, \varphi) ,
\end{array}
\label{eq.2.3.1.1}
\end{equation}
where the nuclear component $V_{\rm N}$ and the Coulomb component $V_{\rm C}$ are
\begin{equation}
\begin{array}{lllccl}
\vspace{2.5mm}
  V_{\rm N}(\mathbf{r}, R_{12}, \varphi) =
    V_{\rm N,\, nucleus}(\mathbf{r}, R_{12}, \varphi) - E_{\rm N,\, fragment}, \\

  V_{\rm C}(\mathbf{r}, R_{12}, \varphi) =
    V_{\rm C,\, nucleus}(\mathbf{r}, R_{12}, \varphi) - E_{\rm C,\, fragment},
\end{array}
\label{eq.2.3.1.2}
\end{equation}
with $V_{\rm N,\, nucleus}(\mathbf{r}, R_{12}, \varphi)$ and $V_{\rm C,\, nucleus}(\mathbf{r}, R_{12}, \varphi)$ the nuclear and Coulomb energies, respectively, of the full nuclear system, while $E_{\rm N,\, fragment}(\mathbf{r})$ and $E_{\rm C,\, fragment}(\mathbf{r})$ are the nuclear and Coulomb energies, respectively, for the separated fragments. They are defined as
\begin{equation}
\begin{array}{lll}
\vspace{1.5mm}
  V_{\rm N,\, nucleus}(\mathbf{r}, R_{12}, \varphi) =
    -\lambda_{N} 
  \hspace{-3.0mm}
  \displaystyle\int\limits_{V (R_{12}, |\mathbf{r}| ,\varphi)}

  \hspace{-1.5mm}
\displaystyle\frac{\mathbf{dr^{\prime}}^{3}}{1 + \exp(|\mathbf{r} - \mathbf{r}^{\prime}|/a)}, \\

\vspace{1.5mm}
  E_{\rm N,\, fragment} =
    -\lambda_{N}\;
    \displaystyle\int\limits_{V}
\displaystyle\frac{\mathbf{dr}^{3}}{1 + \exp(\mathbf{|r|/a})}, \\

\vspace{1.5mm}
  V_{\rm C,\, nucleus}(\mathbf{r}, R_{12}, \varphi) =
    \lambda_{C}\, \displaystyle\int\limits_{V\,(R_{12}, |\mathbf{r}| ,\varphi),\, \mathbf{r} \ne \mathbf{r^{\prime}}}
\displaystyle\frac{\mathbf{dr^{\prime}}^{3}}{|\mathbf{r}^{\prime} - \mathbf{r}|}, \\

  E_{\rm C,\, fragment} =
    \lambda_{C}\, \displaystyle\int\limits_{V,\, r \ne 0}
      \displaystyle\frac{\mathbf{dr}^{3}}{|\mathbf{r}|}.
\end{array}
\label{eq.2.3.1.3}
\end{equation}
Here, we integrate over the volume with the surface of the full nuclear system. In Eq.~(\ref{eq.2.2.1.3}) the surface is arbitrary, 
while in Eq.~(\ref{eq.2.3.1.3}) we already define it on the basis of the three parameters $R_{12}$, $\varphi$ and $|\mathbf{r}|$. Such an assumption of the nuclear surface gives the possibility to calculate the potential.

%

The procedure above allows to calculate the wave functions and bremsstrahlung probabilities. However, in realization of such a method we have met with the following problem. 
(1) Time of calculations of the bremsstrahlung spectra is quite long.
But, 
in order to construct the working model, 
we need to analyze many details of the model, algorithms of calculations, proper choice of parameters, different geometries of separation of binary nucleus, etc. In this first step, we need to obtain the first estimations of the spectra in fast way many times, i.e. we need in fast calculations of the spectra (note that there is no any information about bremsstrahlung for this process, which could be used by us as orientation how to construct the model). 
By such a reason, use of more sophisticated models of ternary fission themselves is not effective in this step of development.
(2) It is difficult to obtain convergent calculation and stable spectra as we have to take into account far distances where we calculate wave function starting from the corresponding potential values.

\begin{figure}[htbp]
\centerline{\includegraphics[width=61mm]{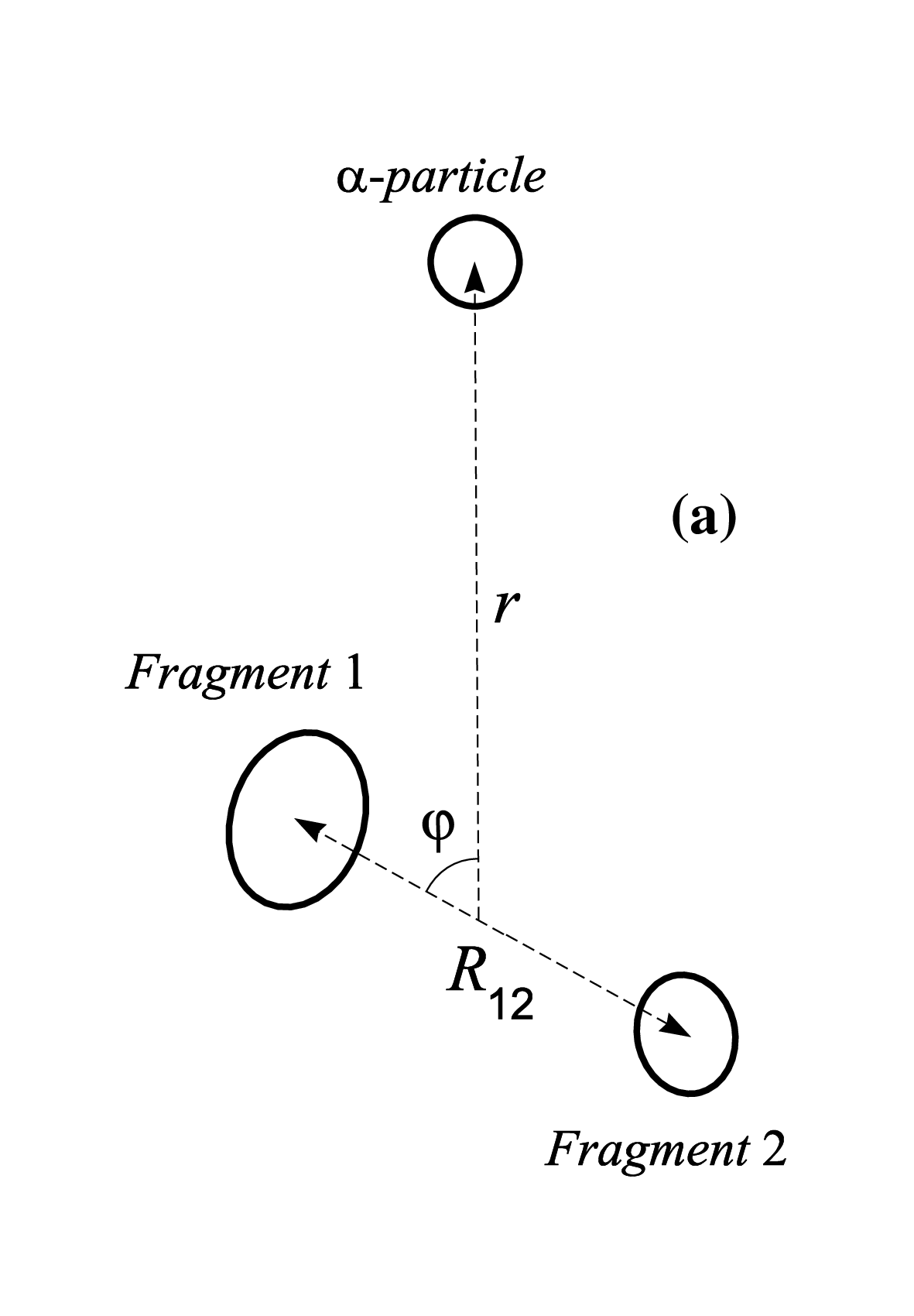}}
\vspace{-18mm}
\centerline{\includegraphics[width=61mm]{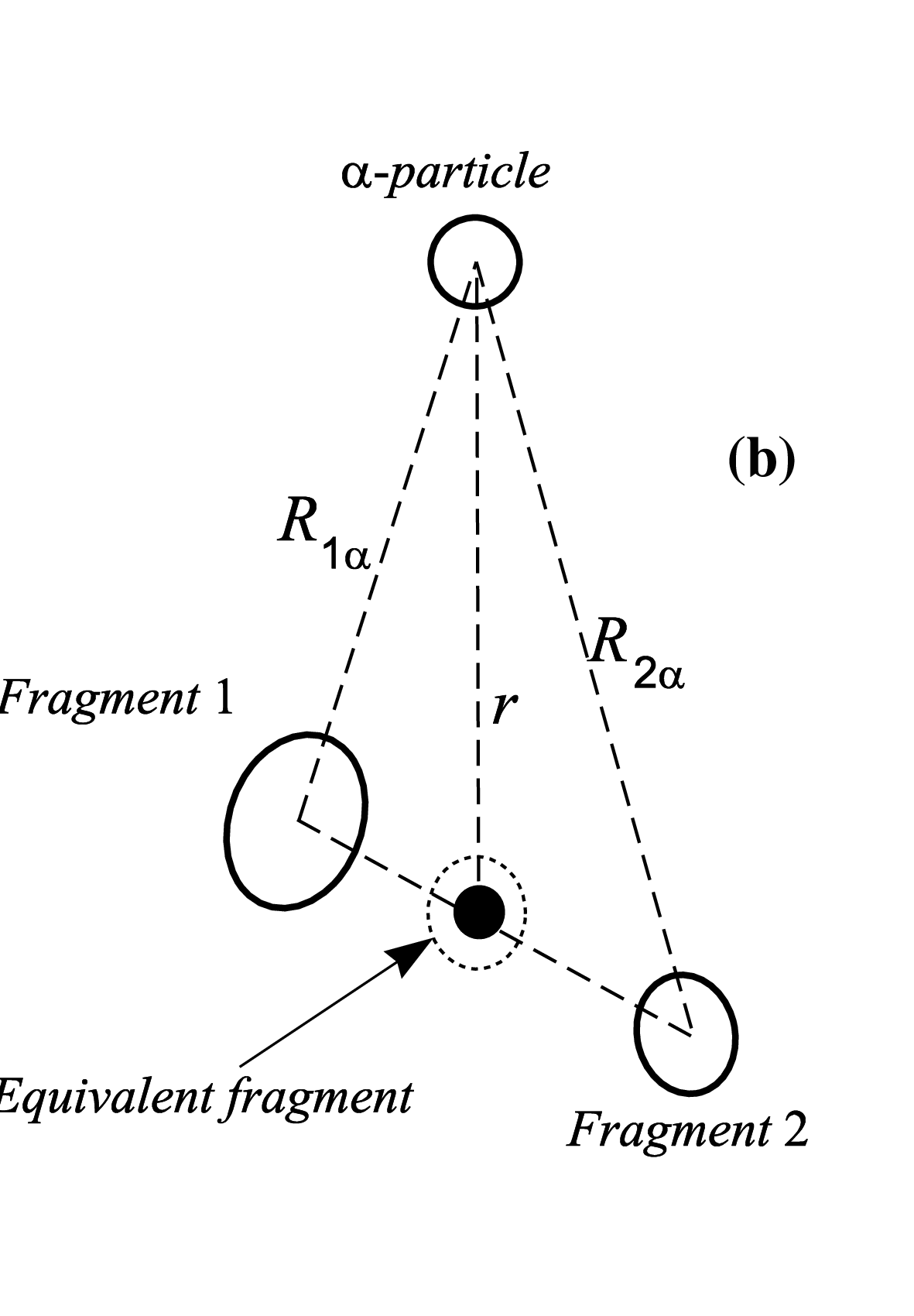}}
\vspace{-10.5mm}
\caption{\small Geometry of fissioning nuclear system at far distances: (a)  for the general geometrical location of the two fragments of the binary nucleus relatively to the $\alpha$\,particle we use only three parameters needed for calculation of the potential:  the $r$ distance between the centers of masses of the $\alpha$\,particle and the binary nucleus, the $R_{\rm 12}$ distance between the centers of masses of the two fragments of the binary nucleus, and the angle  $\varphi$ between the  direction of  separation into two fragments of the binary nucleus and the direction of the $\alpha$ particle motion with respect to the center-of-mass of the daughter binary nucleus; (b) passage to the equivalent fragment with the electromagnetic charge $Z_{\rm eq}$ defined in Eq.~(\ref{eq.2.3.2.1}). \label{fig.2}}
\end{figure}

For this reason, in order to calculate the Coulomb potential component at far distances we substitute the two fragments (located at the distances $\mathbf{R}_{1\alpha}$ and $\mathbf{R}_{2\alpha}$ from the $\alpha$\,particle) with a new fictitious fragment with charge $Z_{\rm eq}$ located in the center-of-mass of the binary nucleus (at the distance $r$ from the $\alpha$\,particle) determining the same influence on the $\alpha$\,particle as the two original fragments (see Fig.~\ref{fig.2}~(b)).
The unknown charge $Z_{\rm eq}$ is defined as
\begin{equation}
  Z_{\rm eq}(r) =
  r\, \Bigl\{ \displaystyle\frac{Z_{1}}{R_{1\alpha}} + \displaystyle\frac{Z_{2}}{R_{2\alpha}} \Bigr\},
\label{eq.2.3.2.1}
\end{equation}
where $r = |\mathbf{r}|$, $R_{1\alpha} = |\mathbf{R}_{1\alpha}|$, $R_{2\alpha} = |\mathbf{R}_{2\alpha}|$. It turns out that such an approach provides stability in calculation of the bremsstrahlung spectra. Another important result 
of this approach is that the bremsstrahlung spectra 
are sensitive to the geometry
of the fissioning nucleus (i.e. $R_{12}$, $r$ and $\varphi$).

It turns out that the potential calculated by this way depends on ratio between $R_{12}$ and $r$. The condition of Eq.~(\ref{eq.2.3.2.1}) defines the electric charge $Z_{\rm eq}$ on the basis of $R_{12}$ and $r$. Therefore, it cannot impose any restriction on the choice of the $R_{12}$ and $r$ values. The calculations of the bremsstrahlung spectra are sensitive to such a dynamical aspect and the chosen fission geometry. We assume a linear relation between $R_{12}$ and $r$ as
\begin{equation}
  R_{12} (t) = R_{12,\, 0} + C_{1}\, r (t),
\label{eq.2.3.3.1}
\end{equation}
where $C_{1}$ is a coefficient, $R_{12,\, 0}$ is the distance between the centers-of-masses of two fragments of the binary nucleus at moment of separation of the $\alpha$\,particle, and $t$ is time. From Eq.~(\ref{eq.2.3.3.1}) the following initial condition can be written down
(we define $t=0$ at $r\, (0) = 0$ as initial time of the process):
\begin{equation}
\begin{array}{lll}
  r\, (0) = 0 \,\,\,\, \mbox{and } &
  R_{12} (0) = R_{12,\, 0} &
  \,\,\mbox{at }\,\, t = 0.
\end{array}
\label{eq.2.3.3.2}
\end{equation}

\subsection{Matrix element of bremsstrahlung  and contribution from the relative motion of the $\alpha$-particle and the binary nucleus
\label{sec.2.4}}

Now we will determine the matrix element of bremsstrahlung taking into account separation of the binary nucleus on two fragments. 
Let's number nucleons for the $\alpha$\,particle by index $i$, and nucleons for the nucleus by index $j$. The operator of the photon emission from the nuclear system (composed of the $\alpha$\,particle and binary nucleus) where the $\alpha$\,particle and residual nucleus are composed of nucleons, is written down as
\begin{equation}
\begin{array}{lcl}
  \hat{H}_{\gamma} =
    -\,e\, \sqrt{\displaystyle\frac{2\pi}{w}}\,
    \displaystyle\sum\limits_{\eta=1,2} \mathbf{e}^{(\eta),*}\;
    \biggl\{
      \displaystyle\sum\limits_{i=1}^{4}
        \displaystyle\frac{Z_{i}}{m_{i}}\; e^{-i \mathbf{k\cdot r}_{i}}\, \mathbf{\hat{p}}_{i} + \\
     ~~~~~~~~~~ \displaystyle\sum\limits_{j=1}^{A}
        \displaystyle\frac{Z_{j}}{m_{j}}\; e^{-i \mathbf{k\cdot r}_{j}}\, \mathbf{\hat{p}}_{j}
    \Bigr\}.
\end{array}
\label{eq.2.4.1}
\end{equation}
Here, $\mathbf{e}^{(\eta)}$ are unit vectors of polarization of the photon, 
$\mathbf{k}$ is wave vector of the photon and $w = k = \bigl| \mathbf{k}\bigr|$. The vectors $\mathbf{e}^{(\eta)}$ are perpendicular to $\mathbf{k}$ in the Coulomb gauge.
We have two independent polarizations $\mathbf{e}^{(1)}$ and $\mathbf{e}^{(2)}$ for the photon with momentum $\mathbf{k}$ ($\eta=1,2$). We use the system of units where $\hbar = c = 1$.
$\mathbf{r}_{i}$ and $\mathbf{r}_{j}$ are the radius-vectors marking  the position of the nucleons of the $\alpha$\,particle and binary nucleus, respectively,  $Z_{i}$ and $Z_{j}$ are 
the electric charges of nucleons of the $\alpha$\,particle and binary nucleus, respectively.

We denote coordinate of the center-of-mass of the $\alpha$\,particle as $\mathbf{r}_{\alpha}$, 
the one of the binary nucleus as $\mathbf{R}_{A}$, and the one for the full nuclear system as $\mathbf{R}$
\begin{equation}
\begin{array}{lll}
\vspace{1.5mm}
\mathbf{r}_{\alpha} = \displaystyle\frac{1}{m_{\alpha}} \displaystyle\sum_{i=1}^{4} m_{i}\, \mathbf{r}_{\alpha i}, &
  \mathbf{R}_{A}      = \displaystyle\frac{1}{M} \displaystyle\sum_{j=1}^{A} m_{j}\, \mathbf{r}_{A j}, \\
  \mathbf{R}          = \displaystyle\frac{M\mathbf{R}_{A} + m_{\alpha}\mathbf{r}_{\alpha}}{M+m_{\alpha}},
\end{array}
\label{eq.2.4.2}
\end{equation}
where $M$ and $m_{\alpha}$ are the masses of the binary nucleus and  $\alpha$\,particle, respectively.
Introducing the  new relative coordinates $\mathbf{s}_{\alpha i}$, $\mathbf{s}_{A j}$ and $\mathbf{r}$
\begin{equation}
\begin{array}{lll}
  \mathbf{r}_{i} = \mathbf{r}_{\alpha} + \mathbf{s}_{\alpha i}, &
  \mathbf{r}_{j} = \mathbf{R}_{A} + \mathbf{s}_{A j}, &
  \mathbf{r} = \mathbf{r}_{\alpha} - \mathbf{R}_{A},
\end{array}
\label{eq.2.4.3}
\end{equation}
we obtain
\begin{equation}
\begin{array}{ll}
  \mathbf{r}_{i} = \mathbf{R} + \displaystyle\frac{M}{M+m_{\alpha}}\: \mathbf{r} + \mathbf{s}_{\alpha i}, \\
  \mathbf{r}_{j} = \mathbf{R} - \displaystyle\frac{m_{\alpha}}{M+m_{\alpha}}\: \mathbf{r} + \mathbf{s}_{A j}.
\end{array}
\label{eq.2.4.4}
\end{equation}
Using such coordinates above, we calculate the matrix element as 
(see Appendix~\ref{sec.app.1}, for details)
%
\begin{equation}
\begin{array}{llll}
\vspace{1.5mm}
  F_{fi} = \langle f |\, \hat{H}_{\gamma} |\,i \rangle =
  -\,e\, \sqrt{\displaystyle\frac{2\pi}{w}}\,   \displaystyle\sum\limits_{\eta=1,2} \mathbf{e}^{(\eta),*}\;
    \delta(\mathbf{K}_{f} - \mathbf{k}) \times \\

\vspace{1.5mm}
    \biggl\{
      \biggl\langle f_{A},\, f_{\alpha} \biggl|\,
        Z_{\rm eff}(\mathbf{r})\: e^{-i\,\mathbf{k\cdot r}}\: \mathbf{p}\;
      \biggr|\,
        i_{A},\, i_{\alpha}
      \biggr\rangle\; + \\
    \Biggl\langle
      f_{\alpha}
      \Biggl|\,
        e^{i\, \mathbf{k\cdot r}\, \displaystyle\frac{m_{\alpha}}{M+m_{\alpha}}}\,
        \Bigl\langle\, f_{A} \Bigl|\,
          Z_{\rm A}(\mathbf{k})\, \mathbf{p}_{A j}\;
        \Bigr|\, i_{A} \Bigr\rangle\,
      \Biggr|\,
      i_{\alpha}
    \Biggr\rangle
    \Biggr\}.
\end{array}
\label{eq.2.4.5}
\end{equation}
%
Here, the indices $i$ and $f$ denote the initial state (the state before the photon emission) and the final state (after the photon emission),  $|\,s_{A}\rangle$ is the wave function describing internal states of the binary nucleus, $s_{\alpha}$ is the wave function describing the relative motion (with possible tunneling) of the $\alpha$\,particle concerning the binary nucleus.
$Z_{\rm eff}(\mathbf{r})$ and $Z_{\rm A} (\mathbf{k})$ are the effective electric charge of the nuclear system ($\alpha$\,particle and binary nucleus) and the 
electric form factor of the  binary nucleus, respectively, as defined 
in Eqs.~(A.9 - A.12) in Appendix~\ref{sec.app.1}.


\subsection{Probability of photon emission  \label{sec.2.5}}

We define the probability of transition of the fissioning nuclear system per time unit from the state before emission of photon (the initial $i$-state) 
to the state after emission of photon (the final $f$-state) in the given interval $d \nu_{f}$, with emission of photon with momentum inside the given interval $d \nu_{ph}$ as
\begin{equation}
\begin{array}{lll}
\vspace{1.5mm}
  d W =
    2\pi \:|F_{fi}|^{2} \: \delta (w_{f} - w_{i} + w) \cdot d\nu ,
  \\

  d \nu = d\nu_{f} \cdot d\nu_{ph},
  \:\:
  d \nu_{ph} = \displaystyle\frac{d^{3} k}{(2\pi)^{3}} =
  \displaystyle\frac{w^{2}\, dw \,d\Omega_{ph}}{(2\pi c)^{3}},
\end{array}
\label{eq.2.5.1}
\end{equation}
where $d\nu_{ph}$ and $d\nu_{f}$ are intervals for
the photon and particle in the final $f$-state, $d\Omega_{ph} = d\,(\cos{\theta_{ph}}) = \sin{\theta_{ph}} \,d\theta_{ph} \,d\varphi_{ph}$, $k_{ph}=w/c$.

The matrix element of Eq.~(\ref{eq.2.4.5}) is rewritten via new variables as
%
\begin{equation}
\begin{array}{ll}
\vspace{1.5mm}
  F_{fi} =
  \displaystyle\frac{e}{m}\,
\sqrt{\displaystyle\frac{2\pi}{w}} \; \tilde{p}\,(k_{i},k_{f}), \\

  \tilde{p}\,(k_{i}, k_{f}) =
    \displaystyle\sum\limits_{\eta=1,2} \mathbf{e}^{(\eta),*}\, \tilde{\mathbf{p}}\,(k_{i}, k_{f}),
\end{array}
\label{eq.2.5.2}
\end{equation}

\noindent
\vspace{-5mm}
\begin{equation}
\begin{array}{lcl}
\vspace{1.5mm}
  \tilde{\mathbf{p}}\,(k_{i}, k_{f}) =
  \biggl\langle f_{A},\, f_{\alpha} \biggl|\,
        Z_{\rm eff}(\mathbf{r})\: e^{-i\,\mathbf{k\cdot r}}\: \mathbf{p}\;
      \biggr|\,
        i_{A},\, i_{\alpha}
    \biggr\rangle\; + \\

  \Biggl\langle
      f_{\alpha}
      \Biggl|\,
        e^{i\, \mathbf{k\cdot r}\, \displaystyle\frac{m_{\alpha}}{M+m_{\alpha}}}\,
        \Bigl\langle\, f_{A} \Bigl|\,
          Z_{\rm A}(\mathbf{k})\, \mathbf{p}_{A j}\;
        \Bigr|\, i_{A} \Bigr\rangle\,
      \Biggr|\,
    i_{\alpha}
  \Biggr\rangle.
\end{array}
\label{eq.2.5.3}
\end{equation}
%
Integrating Eq.~(\ref{eq.2.5.1}) over $dw$ and using Eq.~(\ref{eq.2.5.2}) for $F_{fi}$, we find
\begin{equation}
\begin{array}{cc}
\vspace{1.5mm}
  d W = \displaystyle\frac{e^{2}}{m^{2}}\:
  \displaystyle\frac{w_{fi}}{2\pi}\;
  \Bigl|\tilde{p}(k_{i}, k_{f})\Bigr|^{2}\;
  d\Omega_{ph} \, d\nu_{f}, \\
  w_{fi} = w_{i} - w_{f} = E_{i} - E_{f},
\end{array}
\label{eq.2.5.4}
\end{equation}
where is $k_i=|\mathbf{k}_i|=\sqrt{2\mu E_i}$. This is the probability of the photon emission with the momentum $\mathbf{k}$ (with averaging over the polarizations $\mathbf{e}^{(\eta)}$) where the integration over angles of the particle motion after the photon emission has been performed. To take into account the direction $\mathbf{n}_{\mathbf{r}}^{f}$ of motion (or tunneling) of the particle after emission, we define the previous probability by the following mode: the angular probability concerning the photon emission at angle $\theta$ is the function for which the integral on the angle $\theta$ from 0 to $\pi$ corresponds exactly to the total probability of photon emission
\begin{equation}
\begin{array}{ccl}
  \displaystyle\frac{d W(\theta_{f})} {d\,\Omega_{ph} d\cos{\theta_{f}}} =
  \displaystyle\frac{e^{2}}{\pi} \displaystyle\frac{w_{fi}}{m^{2}}\,
    \Re
    \biggl\{
      \tilde{p} (k_{i},k_{f}) \displaystyle\frac{d\, \tilde{p}^{*}(k_{i},k_{f}, \theta_{f})}{d\cos{\theta_{f}}}
    \biggr\},
\end{array}
\label{eq.2.5.5}
\end{equation}
where $\Re\, (f)$ is real part of function $f$. This probability is inversely proportional to the  normalized volume $V$. 
To obtain the probability independent from $V$, we divide Eq.~(\ref{eq.2.5.5}) on the flux $j$ of the outgoing $\alpha$\,particles which is also inversely proportional to this volume $V$. Using the quantum field theory approach
(where $v(\mathbf{p}) = |\mathbf{p}| / p_{0}$ at $c=1$),
we write
\begin{equation}
\begin{array}{cc}
  j = n_{i}\, v(\mathbf{p}_{i}), &
  v_{i} = |\mathbf{v}_{i}| = \displaystyle\frac{c^{2}\,|\mathbf{p}_{i}|} {E_{i}} =
          \displaystyle\frac{\hbar\,c^{2}\,k_{i}} {E_{i}},
\end{array}
\label{eq.2.5.6}
\end{equation}
where $n_{i}$ is the average number of $\alpha$-particles
per time unit before the photon emission (we have $n_{i}=1$ for the normalized wave function in the initial $i$-state), $v(\mathbf{p}_{i})$ is the module of velocity of outgoing $\alpha$-particles,
we obtain the differential absolute probability
(let's name $dW$ as the relative probability)
\begin{equation}
\begin{array}{llll}
\vspace{2.0mm}
  \displaystyle\frac{d\,P (\varphi_{f}, \theta_{f})}{d\Omega_{ph}\, d\cos{\theta_{f}}} \, = \,
  \displaystyle\frac{d\,W (\varphi_{f}, \theta_{f})}{d\Omega_{ph}\, d\cos{\theta_{f}}} \,
    \displaystyle\frac{E_{i}} {\hbar\, c^{2}\, k_{i}} = \\
    
\displaystyle\frac{e^{2}}{\pi}\:
   \displaystyle\frac{w_{ph}\,E_{i}}{m^{2}\,k_{i}} \;
      \,\Re\, \biggl\{\tilde{p}\,(k_{i},k_{f})
        \displaystyle\frac{d\, \tilde{p}^{*}(k_{i},k_{f}, \Omega_{f})}{d\,\cos{\theta_{f}}} \biggr\}.
\end{array}
\label{eq.2.5.7}
\end{equation}

The matrix element of Eq.~(\ref{eq.2.5.3}) is rewritten
as
\begin{equation}
  \tilde{p}\,(k_{i}, k_{f}) =
  \tilde{p}_{\alpha}\,(k_{i}, k_{f}) + \tilde{p}_{\rm binary}\,(k_{i}, k_{f}),
\label{eq.2.5.8}
\end{equation}
with
\begin{widetext}
\begin{equation}
\begin{array}{lcl}
  \tilde{p}_{\alpha}\,(k_{i}, k_{f}) & = &
    \displaystyle\sum\limits_{\eta=1,2} \mathbf{e}^{(\eta),*}\,
  \biggl\langle f_{A},\, f_{\alpha} \biggl|\,
        Z_{\rm eff}(\mathbf{r})\: e^{-i\,\mathbf{k\cdot r}}\: \mathbf{p}\;
      \biggr|\,
        i_{A},\, i_{\alpha}
    \biggr\rangle, \\
  \tilde{p}_{\rm binary}\,(k_{i}, k_{f}) & = &
    \displaystyle\sum\limits_{\eta=1,2} \mathbf{e}^{(\eta),*}\,
    \Biggl\langle
      f_{\alpha}
      \Biggl|\,
        e^{i\, \mathbf{k\cdot r}\, \displaystyle\frac{m_{\alpha}}{M+m_{\alpha}}}\,
        \Bigl\langle\, f_{A} \Bigl|\,
          Z_{\rm A}(\mathbf{k})\, \mathbf{p}_{A j}\;
        \Bigr|\, i_{A} \Bigr\rangle\,
      \Biggr|\,
    i_{\alpha}
  \Biggr\rangle.
\end{array}
\label{eq.2.5.9}
\end{equation}
\end{widetext}
The first term in
Eq. (\ref{eq.2.5.9}) is calculated further as
\begin{equation}
\begin{array}{cclccl}
\vspace{1.5mm}
\tilde{p}_{\alpha}(w, \vartheta) & = &
    -\sqrt{\displaystyle\frac{1}{3}}  \displaystyle\sum\limits_{l=0}^{+\infty}
    i^{l} (-1)^{l} \: (2l+1) \: P_{l}(\cos{\vartheta})
    \times \\

  & &
\displaystyle\sum\limits_{\eta = -1, 1} h_{\eta} J_{m_{f}}(l,w),
%
\end{array}
\label{eq.2.5.10}
\end{equation}
Here, 
$h_{\pm} = \mp (1 \pm i) / \sqrt{2}$, 
$\mathbf{k}$ is the momentum of the photon, $\mathbf{r}$ is radius-vector marking the position of the center-of-mass of the $\alpha$\,particle relatively to the center-of-mass of the binary nucleus, $\theta$ is the angle between the vectors $\mathbf{n}_{r} = \mathbf{r}/r$ and $\mathbf{n}_{ph} = \mathbf{k}/k$, 
$k = |\mathbf{k}|$ and $r = |\mathbf{r}|$.
$E_{i,f}$ and $k_{i,f}$ are the total energy and wave vector of the system, respectively, in the initial $i$-state or the final $f$-state, where $w = k = \bigl|\mathbf{k}\bigr|$ (see Refs.~\cite{Maydanyuk:2002ag,Giardina:2008pwx,Giardina:2008sd,Maydanyuk:2022nip}
for details, reference therein).
The radial integral $J_{m_{f}}(l,w)$ is
\begin{equation}
  J_{m_{f}}(l,w) =
    \int\limits^{+\infty}_{0}
    Z_{\rm eff}(r)\; r^{2} \, R^{*}_{f}(r, E_{f}) \,
    \displaystyle\frac{\partial R_{i}(r, E_{i})} {\partial r} \, j_{l} (k\, r) \: dr,
\label{eq.2.5.11}
\end{equation}
where $R_{i}(r)$ and $R_{f}(r)$ are the radial parts of wave function $\psi_{i}(\mathbf{r})$ of the nuclear system at the initial $i$-state and the wave function $\psi_{f}(\mathbf{r})$ of the nuclear system at the final $f$-state,
$j_{l}(kr)$ is the spherical Bessel function of the order $l$.

\subsection{Contribution from relative motion of heavy fragments of the binary nucleus 
\label{sec.2.binary}}

Let's rewrite the total matrix element of photon emission as sum of two terms 
[see Eqs.(\ref{eq.2.5.8}-\ref{eq.2.5.9})] as
%
%
\begin{equation}
  \Bigl\langle f \Bigl| \hat{H}_{\nu} \Bigr| i \Bigr\rangle =
  \Bigl\langle f \Bigl| \hat{H}_{\nu} \Bigr| i \Bigr\rangle_{\alpha} +
  \Bigl\langle f \Bigl| \hat{H}_{\nu} \Bigr| i \Bigr\rangle_{\rm binary}.
\label{eq.5.2.1}
\end{equation}
Here, the second term represents the emission of photons caused by the relative motion of two fragments of the binary nucleus, and it is written down 
as
\begin{widetext}
\begin{equation}
\begin{array}{lcl}
  \Bigl\langle f \Bigl| \hat{H}_{\gamma} \Bigr| i \Bigr\rangle_{\rm binary} & = &
  -\,e\, \sqrt{\displaystyle\frac{2\pi\hbar}{w}}\,
    \displaystyle\sum\limits_{\eta=1,2} \mathbf{e}^{(\eta),*}\;
    \biggl\langle f \biggl|\,
      e^{i\, \mathbf{k\cdot r}\, \displaystyle\frac{m_{\alpha}}{M+m_{\alpha}}} \cdot
      \Bigl\langle f_{A} \Bigl|\, Z_{\rm eff, A} (k, R_{12}, \xi)\; \mathbf{p}_{\rm A_{j}} \Bigr|i_{A} \Bigr\rangle
    \biggr| i \biggr\rangle = \\
  & = &
  -\,e\, \sqrt{\displaystyle\frac{2\pi\hbar}{w}}\,
    \displaystyle\sum\limits_{\eta=1,2} \mathbf{e}^{(\eta),*}\;
    \biggl\langle f \biggl|\,
      e^{i\, \mathbf{k\cdot r}\, \displaystyle\frac{m_{\alpha}}{M+m_{\alpha}}}\,
    \biggr| i \biggr\rangle \cdot
    \Bigl\langle f_{A} \Bigl|\, Z_{\rm eff, A} (k, R_{12}, \xi)\; \mathbf{p}_{\rm A_{j}} \Bigr| i_{A} \Bigr\rangle,
\end{array}
\label{eq.5.2.2}
\end{equation}
where
\begin{equation}
\begin{array}{lcl}
\vspace{2mm}
  Z_{\rm eff, A} (k, R_{12}, \xi) & = &
    e^{i\,k R_{12}\,\cos \xi}\:
    \biggl\{
      e^{- i\,k R_{12}\,\cos\xi\, \displaystyle\frac{m_{1}}{m_{1}+m_{2}}}\:
        \displaystyle\frac{m_{1}\, Z_{2}}{m_{1}+m_{2}} -
      e^{i\,k R_{12}\,\cos\xi\, \displaystyle\frac{m_{2}}{m_{1}+m_{2}}}\:
        \displaystyle\frac{m_{2}\, Z_{1}}{m_{1}+m_{2}}
    \biggr\} \simeq \\
  & \simeq &
    \displaystyle\frac{m_{1}\, Z_{2} - m_{2}\, Z_{1}}{m_{1}+m_{2}}.
\end{array}
\label{eq.5.2.2.1}
\end{equation}
\end{widetext}
Here, $k=|\mathbf{k}|$,
$\xi$ is angle between vectors $\mathbf{k}$ and $\mathbf{R}_{12}$,
$m_{1}$ and $m_{2}$ are masses of two fragments of the binary nucleus, $Z_{1}$ and $Z_{2}$ are electric charges of these
fragments (we have $m_{1}+m_{2}=M$, $Z_{1}+Z_{2}=Z_{\rm A}$).
It turns out that calculation of the first matrix element
$\Bigl\langle f \Bigl|\,
  \exp\bigl[i\, \mathbf{k\cdot r}\, \displaystyle\frac{m_{\alpha}}{M+m_{\alpha}} \bigr]\,
\Bigr| i \Bigr\rangle$
in Eqs.~(\ref{eq.5.2.2}) is a difficult task,
which is highly important in calculation of the full spectrum of bremsstrahlung photons.


\section{Numerical results \label{sec.3}}

\begin{figure}[htbp]
\centerline{\includegraphics[width=88mm]{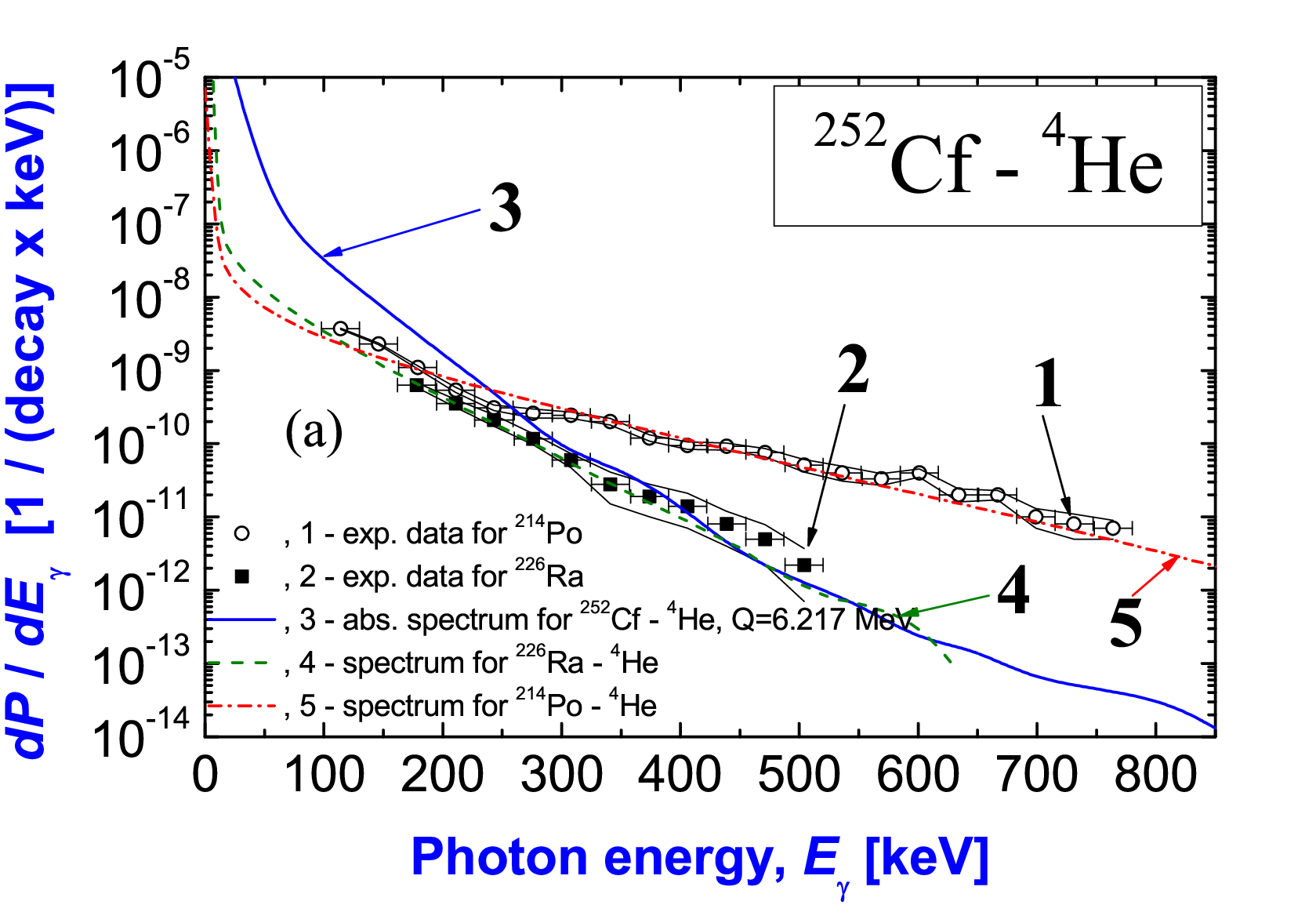}}
\vspace{-2.0mm}
\centerline{\includegraphics[width=88mm]{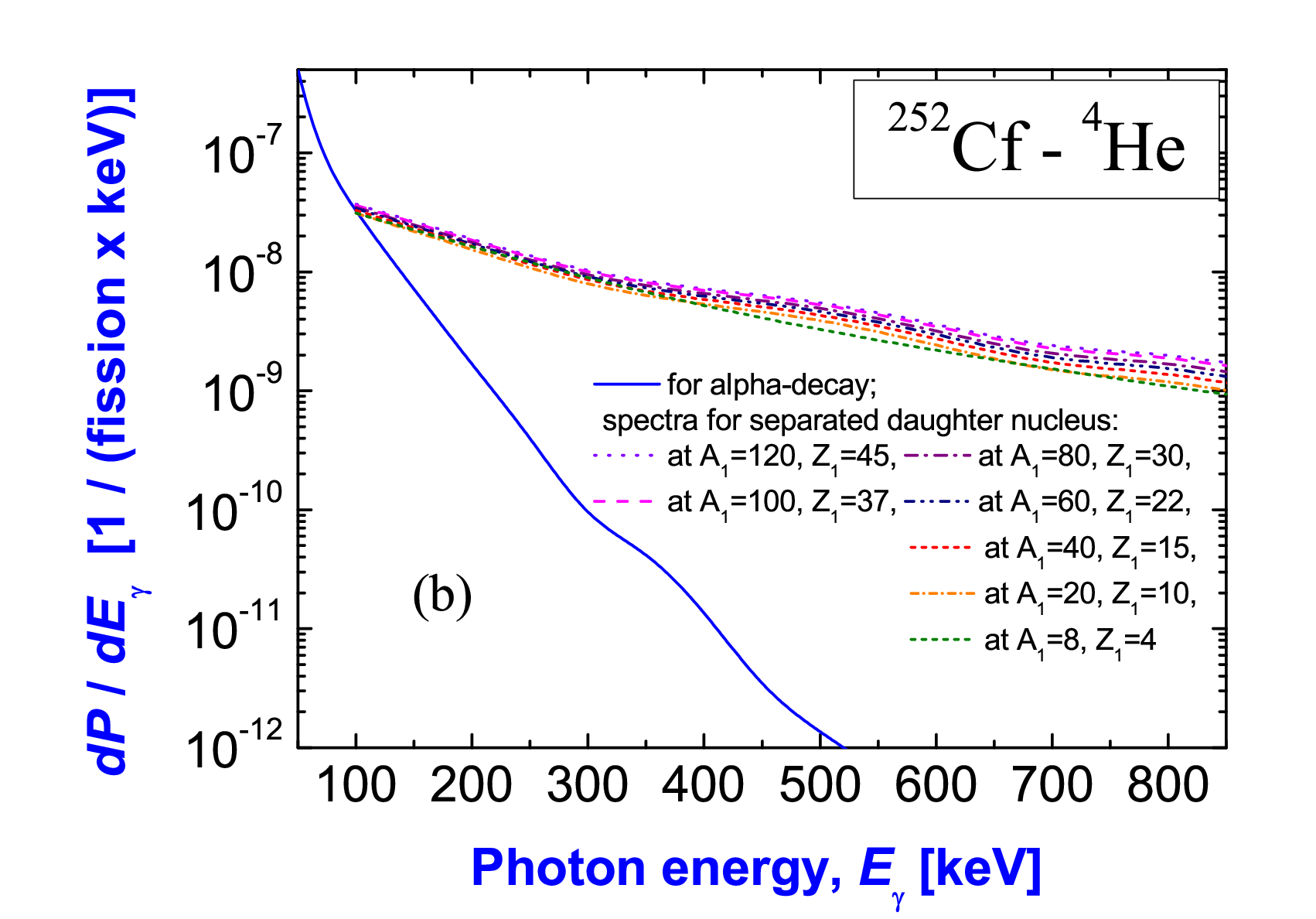}}
\vspace{-4mm}
\caption{\small (Color online)
Bremsstrahlung emission probabilities accompanying the $\alpha$ emission from the \isotope[252]{Cf} nucleus.
Full line represents the photon spectrum caused by emission of the $\alpha$\,particle ($E_{\alpha}$=6.217 MeV) when the binary nucleus is not separated on two fragments. 
The other lines represent the photon spectra caused by the emission of the  $\alpha$\,particle ($E_{\alpha}\simeq$ 16 MeV) taking into account the separation of the binary nucleus on two fragments with different masses.
For the others parameters we choose $R_{12,\,0}=7$~fm, $\varphi=90^{\circ}$, $C_{1}=0.3$.
\label{fig.3}}
\end{figure}

We start from analysis of the bremsstrahlung photons emitted by the $\alpha$\,particle leaving the \isotope[252]{Cf} nucleus when the binary nucleus ($^{248}{\rm Cm}$) is not separated.
The probability of such bremsstrahlung emission is shown in Fig. \ref{fig.3}~(a) by the blue line. One can see that this spectrum is enough close to the bremsstrahlung spectra in the  $\alpha$\,decay of the 
\isotope[214]{Po} and \isotope[226]{Ra} nuclei \cite{Giardina:2008pwx,Giardina:2008sd}.
This result confirms the definition in Eq.~(\ref{eq.2.5.7}) of the probability of photon emission in the absolute scale. In Fig. \ref{fig.3}~(b) we present a new calculation of the probability of photons emitted by the $\alpha$\,particle leaving the \isotope[252]{Cf} nucleus 
when the binary nucleus is separated on two fragments. In the calculation we use energy of the $\alpha$\,particle, $E_{\alpha}=16$~MeV, as in Ref.~\cite{Hwang:2000nt}. One can see that the difference between these spectra and the spectrum for $\alpha$ decay (blue line) is significant. This result is explained by the higher $E_{\alpha}$ energy of the emitted $\alpha$\,particle in case of separation of the binary nucleus on two fragments.

%

\begin{figure}[htbp]
\centerline{\includegraphics[width=88mm]{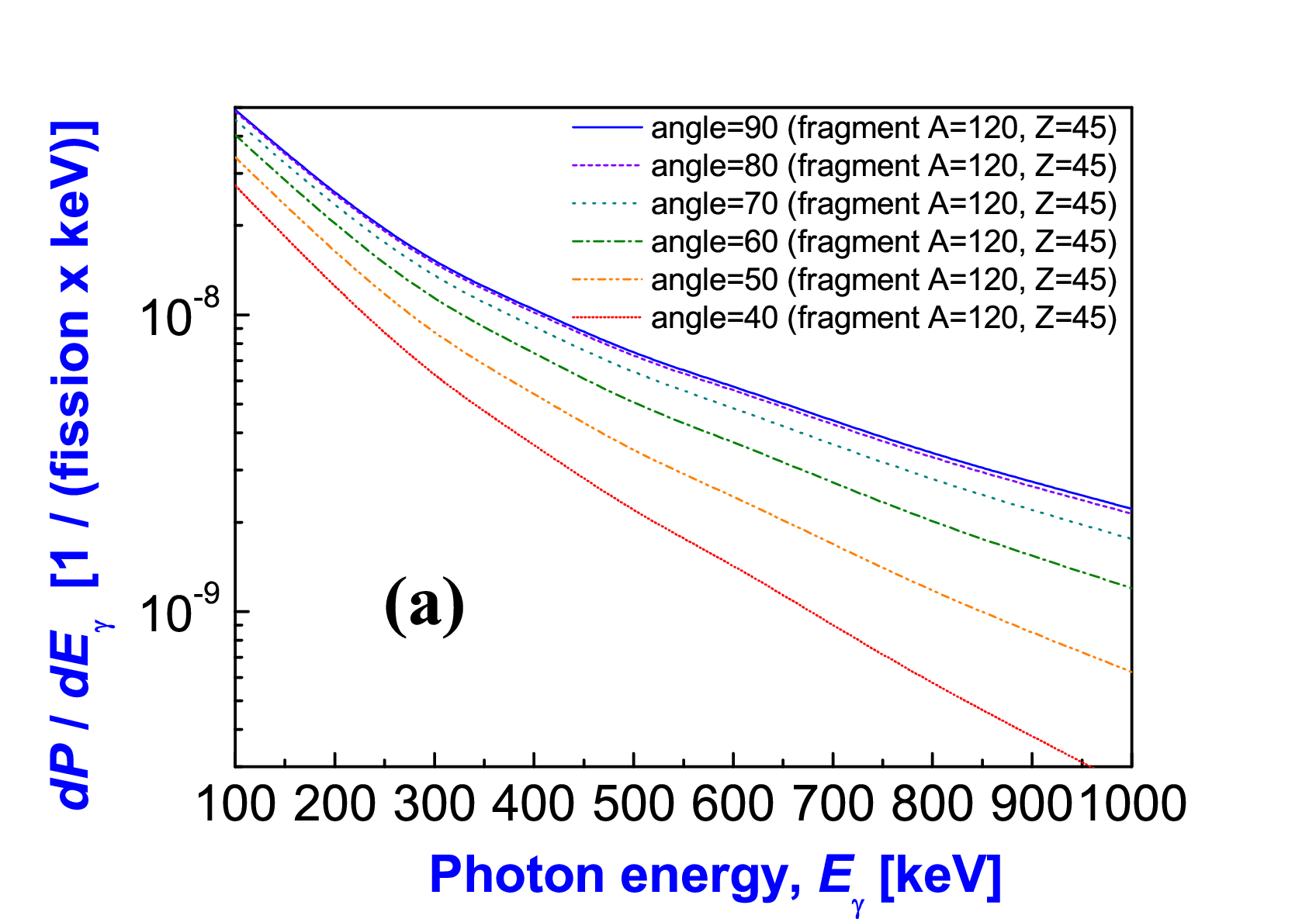}}
\centerline{\includegraphics[width=88mm]{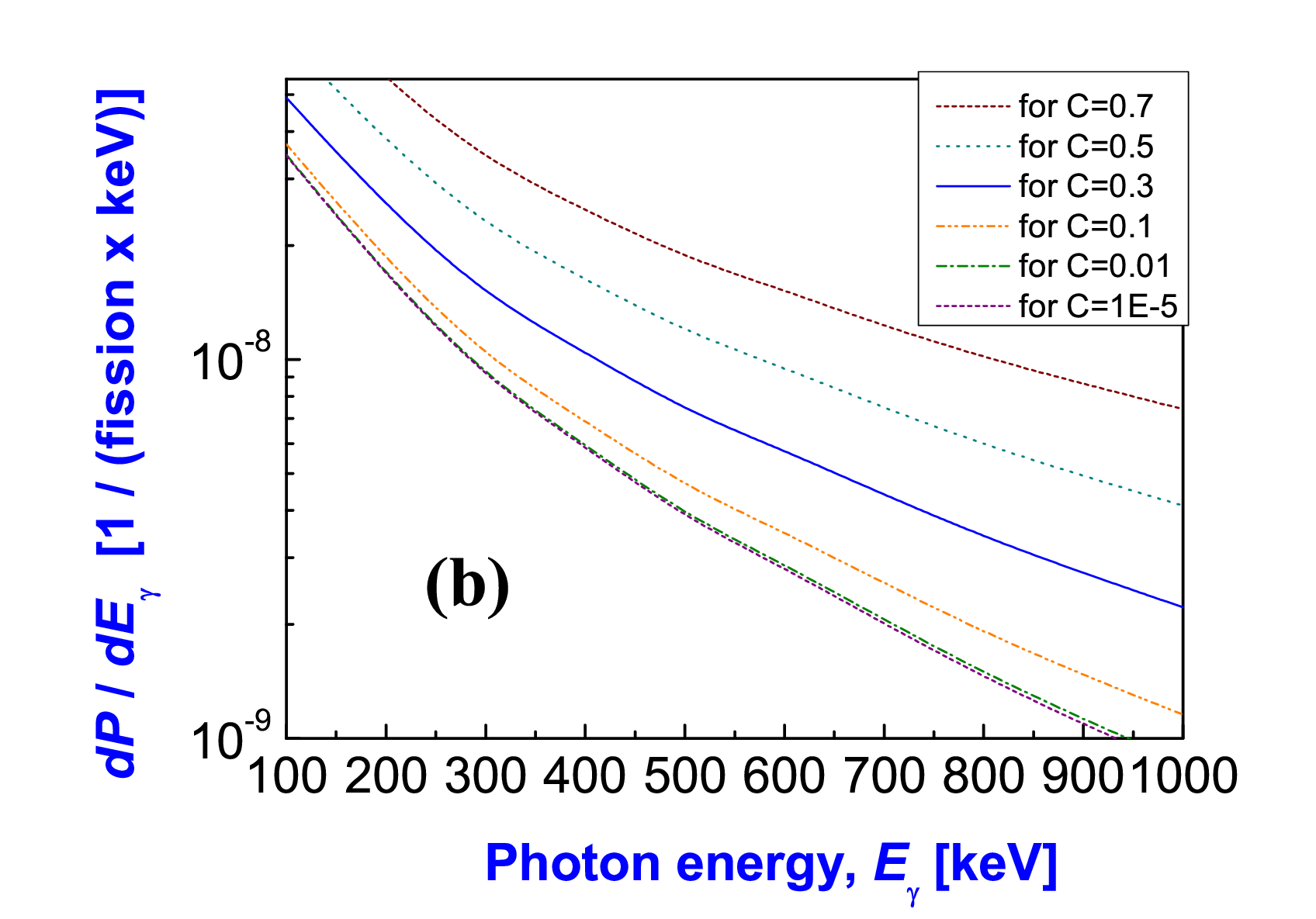}}
\vspace{-4mm}
\caption{\small (Color online)
The bremsstrahlung probabilities of  emitted photons caused by the $\alpha$\,particle leaving the \isotope[252]{Cf} nucleus when the binary nucleus is separated into two fragments with $A_{1}=120$, $Z_{1}=45$ and $A_{2}=128$, $Z_{2}=51$: for the  $R_{12,\,0}$ parameter we used the same previous value of 7 fm.
(a) The spectra depending on the $\varphi$ angle between the axis connecting the two fragments  and the motion of  the $\alpha$\,particle  with respect to the center-of-mass of the binary nucleus. Full line for $\varphi$=90$^\circ$, short dotted line for $\varphi$=80$^\circ$, dotted line for $\varphi$=70$^\circ$, dash-dotted line for $\varphi$=60$^\circ$, dash-double dotted line for $\varphi$=50$^\circ$, and dashed line for $\varphi$=40$^\circ$. For this calculation we used the same previous $C_{1}$ value of 0.3.
(b) The spectra depending on the chosen $C_{1}$ coefficient and at fixed angle value $\varphi=90^{\circ}$. Short dotted line $C_{1}$=0.7, dotted line $C_{1}$=0.5, full line $C_{1}$=0.3, dash-double dotted line $C_{1}$=0.1, dash-dotted line $C_{1}$=0.01, and dashed line $C_{1}$=10$^{-5}$. One can see that the bremsstrahlung emission is more intensive at larger $C_{1}$.
\label{fig.4}}
\end{figure}

The emission of photons is
dependent on the geometry of relative 
separation and leaving of two fragments and
$\alpha$\,particle. Fig.~\ref{fig.4}~(a) shows the dependence of the spectrum of photons on the $\varphi$ angle. 
For demonstration,
we have chosen the same separation of the binary nucleus on the fragments $A_{1}=120$, $Z_{1}=45$ and $A_{2}=128$, $Z_{2}=51$. 
The parameters $R_{12,\,0}$ and $C_{1}$ are fixed in calculation in Fig.~\ref{fig.4} (we use $R_{12,\,0}=7$~fm and $C_{1}=0.3$).
Result of  Fig. \ref{fig.4}~(a) shows that 
the most intensive photon emission 
occurs
at perpendicular leaving of the $\alpha$\,particle with respect to the 
fission axis.
The direction of the emitted $\alpha$\,particle at $\varphi$=90$^\circ$ is also reported in literature as the most probable in the ternary fission.
%

Next we will explore role of relative motion of two fragments concerning the moving $\alpha$\,particle in the bremsstrahlung emission. 
Such probabilities are shown in  Fig.~\ref{fig.4}~(b)
in dependence of the $C_{1}$ coefficient. 
As in the previous case, we chosen the same separation of the binary nucleus 
on
two fragments with $A_{1}=120$, $Z_{1}=45$ and $A_{2}=128$, $Z_{2}=51$, 
and fix
parameters  $R_{12,\,0}=7$~fm and $\varphi=90^{\circ}$.
Such a result confirms significant enhancement of the bremsstrahlung probability due to relative motion of
two fragments of the binary nucleus. 
This is a new 
result in theory of the bremsstrahlung 
accompanying the nuclear decays and fission.

\begin{figure}[htbp]
\centerline{\includegraphics[width=90mm]{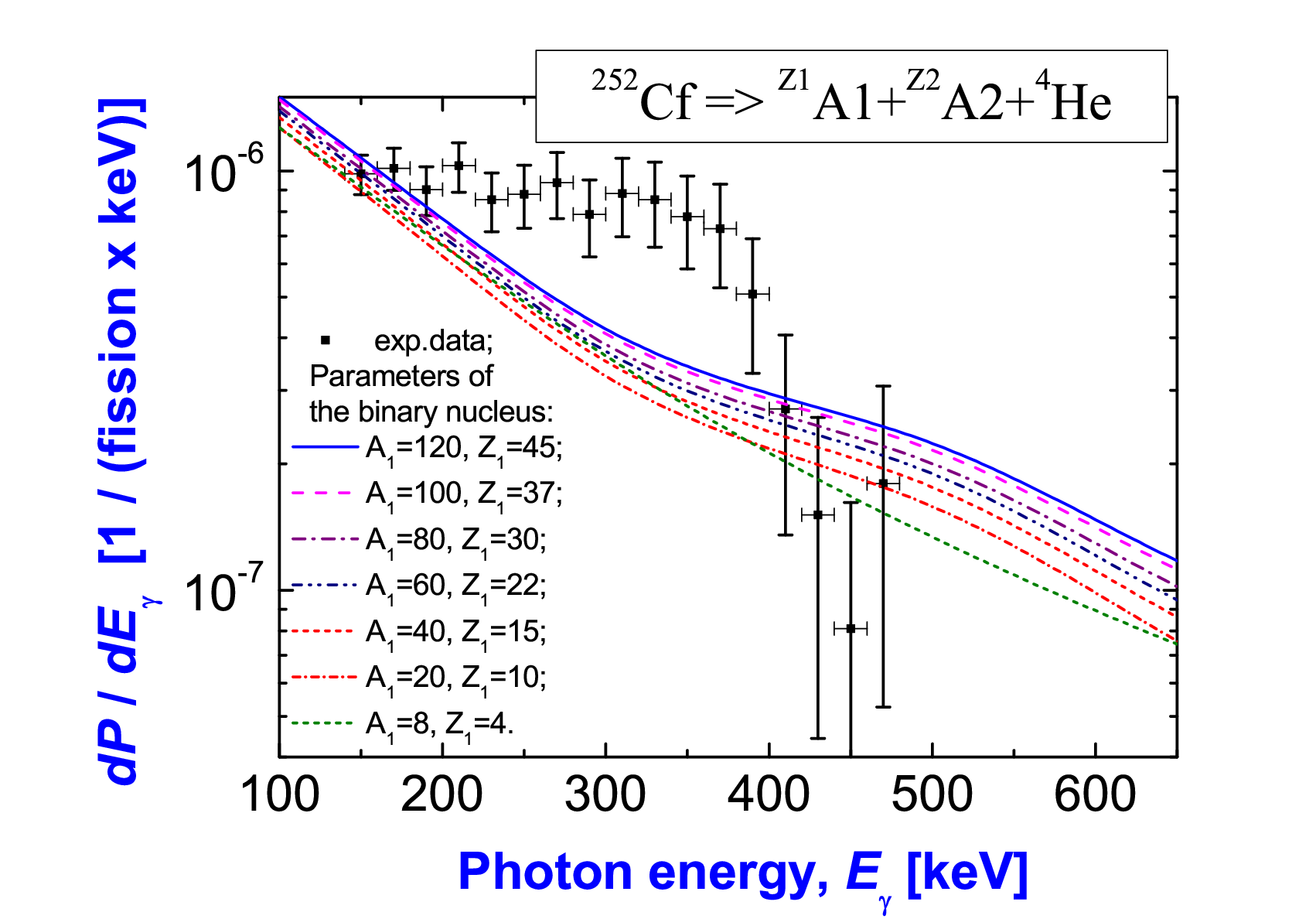}}
\vspace{-3.5mm}
\caption{\small (Color online)
The bremsstrahlung probabilities in  ternary fission of the \isotope[252]{Cf} nucleus obtained for various mass separations of the binary nucleus into two fragments. For the other  parameters we used in calculation $R_{12,\,0}=7$~fm, $\varphi=90^{\circ}$, and $C_{1}=0.3$. Full line for A$_1$=120 (Z$_1$=45) and A$_2$=128 (Z$_2$=51), long dashed line for A$_1$=100 (Z$_1$=37) and A$_2$=148 (Z$_2$=59), dash-dotted line for A$_1$=80 (Z$_1$=30) and A$_2$=168 (Z$_2$=66), dash-double dotted line for A$_1$=60 (Z$_1$=22) and A$_2$=188 (Z$_2$=74), dotted line for A$_1$=40 (Z$_1$=15) and A$_2$=208 (Z$_2$=81), short dashed line for A$_1$=20 (Z$_1$=10) and A$_2$=128 (Z$_2$=86), short dotted  line for A$_1$=8 (Z$_1$=4) and A$_2$=240 (Z$_2$=92).
\label{fig.5}}
\end{figure}

The bremsstrahlung  probabilities calculated for different mass separations of the binary nucleus on two fragments are shown in Fig.~\ref{fig.5}. 
Here, we present the contributions to the full bremsstrahlung spectrum due to the ternary fission of \isotope[252]{Cf} with the $\alpha$ emission and separation of the binary nucleus on two fragments when mass number of one fragment $A_1$ is 120~($Z_1 =45$), 100~($Z_1 =37$), 80~($Z_1 =30$), 60~($Z_1$=22), 40~($Z_1$=15), 20~($Z_1$=10) and 8~($Z_1$=4). 
We used
$R_{12,\,0}=7$~fm, $\varphi=90^{\circ}$, and $C_{1}=0.3$.
This result shows 
that the general tendency of the calculated spectra is similar to the preliminary experimental data in Ref. \cite{Maydanyuk_2011} obtained as a private communication.
There is some disagreement between theory and experiment in the photon energy range of 200--400~keV.
So, it is useful to wait for appearance of the safe experimental bremsstrahlung data for the ternary fission.
In such a situation it is useful to clarify a question: 
how to extract new information about the ternary fission from the analysis of the bremsstrahlung spectra.
To investigate this question, 
we will improve the model with the aim of understanding the sensitive condition of the nuclear system with respect to its dynamics during the ternary fission.

\begin{figure}[htbp]
\centerline{\includegraphics[width=90mm]{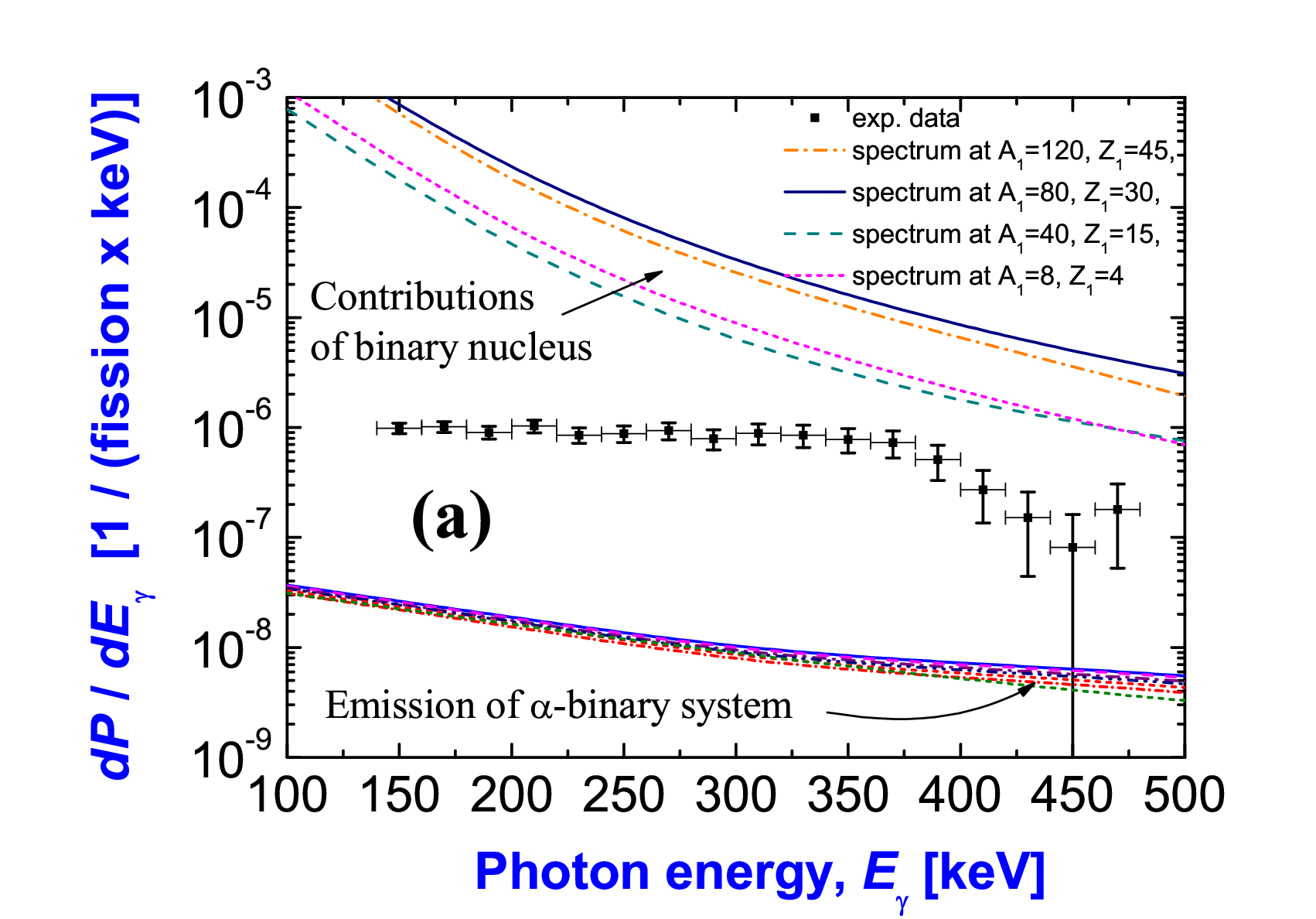}}
\vspace{-3mm}
\caption{\small (Color online)
The bremsstrahlung probabilities formed by the system composed of the $\alpha$\,particle and binary nucleus versus own contributions formed by the two moving fragments of  binary nucleus, calculated for the same chosen mass separation A$_1$~(Z$_1$) and A$_2$~(Z$_2$) as already presented in Fig. \ref{fig.3}. The upper lines represent the photon emission when the $\alpha$\,particle is in presence of the field of the separated fragments of the binary nucleus, while the lower lines represent the photon emission accompanying the $\alpha$\,particle with the unseparated binary nucleus $^{248}$Cm.
The preliminary experimental data~\cite{Maydanyuk_2011} are also added on the figure.
\label{fig.6}}
\end{figure}

It turns out that in calculation of the full bremsstrahlung spectrum the contribution of emitted photons from separation of the binary nucleus on two fragments given
in Eqs.~(\ref{eq.5.2.2})--(\ref{eq.5.2.2.1}) is of the high importance.
A main difficulty in such calculations is to find the matrix element
$\Bigl\langle f \Bigl|\,
  \exp\bigl[i\, \mathbf{k\cdot r}\, \displaystyle\frac{m_{\alpha}}{M+m_{\alpha}} \bigr]\,
\Bigr| i \Bigr\rangle$.
So, we apply the following approximation
\begin{equation}
\begin{array}{lcl}
  \biggl\langle f \biggl|\,
    e^{i\, \mathbf{k\cdot r}\, \displaystyle\frac{m_{\alpha}}{M+m_{\alpha}}}\,
  \biggr| i \biggr\rangle \simeq
  N_{i}^{2} = \displaystyle\frac{\mu}{k_{i}},
\end{array}
\label{eq.5.2.3}
\end{equation}
where $N_{i}$ is the normalization factor for the wave function describing  the state before the emission of photon 
(see Refs.~\cite{Maydanyuk:2002ag,%
Giardina:2008pwx,
Giardina:2008sd
} for details),
$\mu$  is the reduced mass of 
the $\alpha$-particle and the binary nucleus, $\mu = m_{\alpha}\,M /(m_{\alpha} + M)$.
Within this approximation, we find the contributions caused by the motion of fragments of the binary nucleus, based on the matrix element~(\ref{eq.5.2.2}). Calculation for various mass separations of the fragments are shown in Fig.~\ref{fig.6}.

In the next correction to the approximation of Eq.~(\ref{eq.5.2.3}), we take into account dependence of this matrix element both on the wave vector $\mathbf{k}$ of the emitted photon, and the angle between the vectors $\mathbf{k}$ (the direction of the emission of photon) and $\mathbf{r}$ (the direction of motion of $\alpha$\,particle).
Thus, we write down such an approximation as
\begin{equation}
\begin{array}{lcl}
  \biggl\langle f \biggl|\,
    e^{i\, \mathbf{k\cdot r}\, \displaystyle\frac{m_{\alpha}}{M+m_{\alpha}}}\,
  \biggr| i \biggr\rangle \simeq
  \displaystyle\frac{\mu}{k_{i}} \cdot F\,[k, \cos(\mathbf{k}, \mathbf{r})],
\end{array}
\label{eq.5.2.4}
\end{equation}
where $F\,[k, \cos(\mathbf{k}, \mathbf{r})]$ is a new unknown function of the emitted photon energy  and of the angle between the direction of the emitted photon and the motion of the $\alpha$\,particle.

Now let us introduce the following dynamical scheme of the ternary fission.
\begin{description}
\item[a)]
In the starting stage of the ternary fission the $\alpha$\,particle is separated from the full nuclear system, and it moves outside 
and emits photons of low energies. The fragments of the binary nucleus are shifted very slowly, increasing neck. They continue to form the binary nucleus. Thus, the contribution to the photon emission due to this relative motion of fragments exists but it is very small and can be neglected.

\item[b)]At the second stage of the ternary fission, the fragments of the binary nucleus are separated and begin to move outside with acceleration. 
Those produce emission of photons of high energies, which becomes prevailing.
\end{description}
\begin{figure}[htbp]
\centerline{\includegraphics[width=89mm]{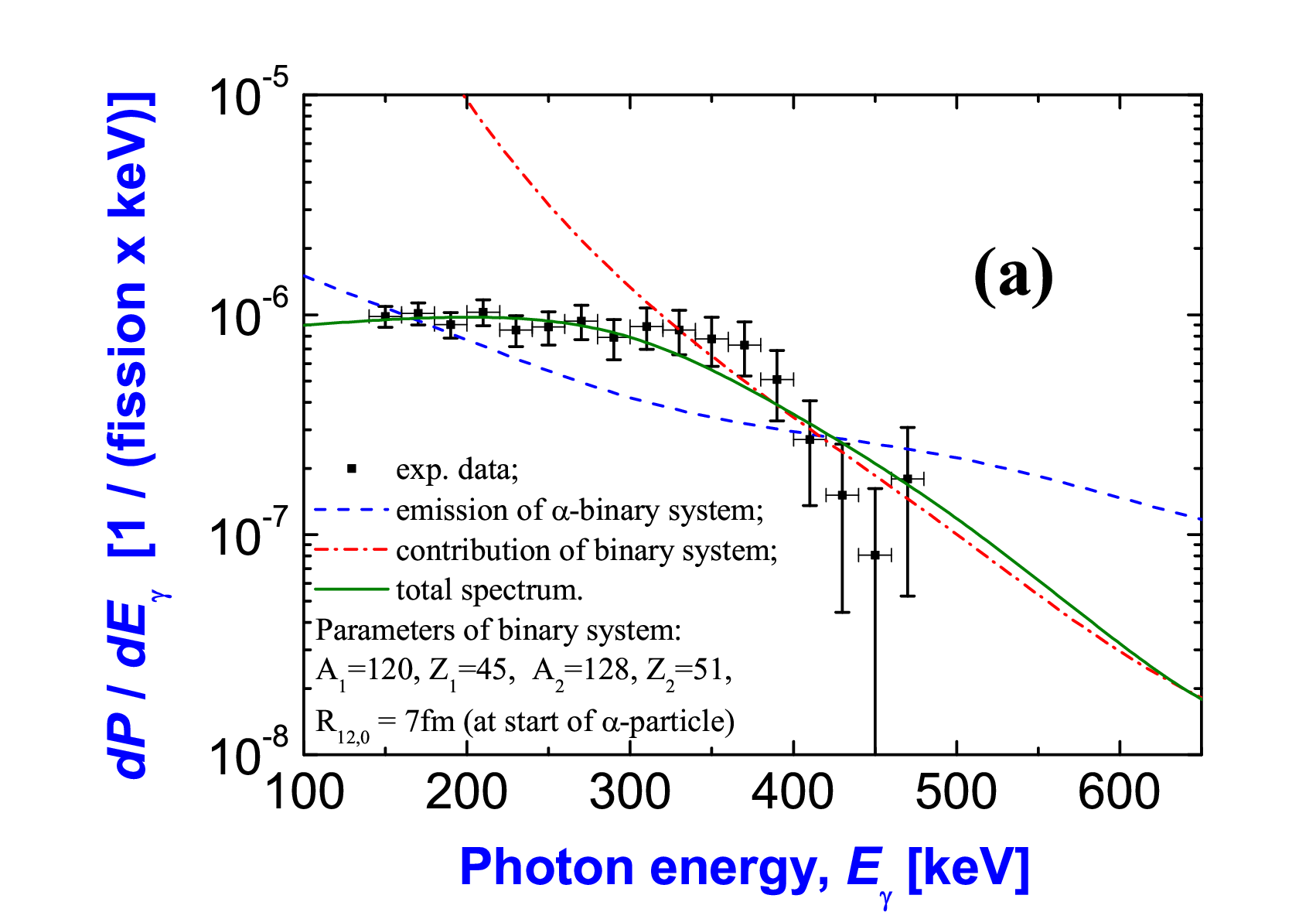}}
\centerline{\includegraphics[width=89mm]{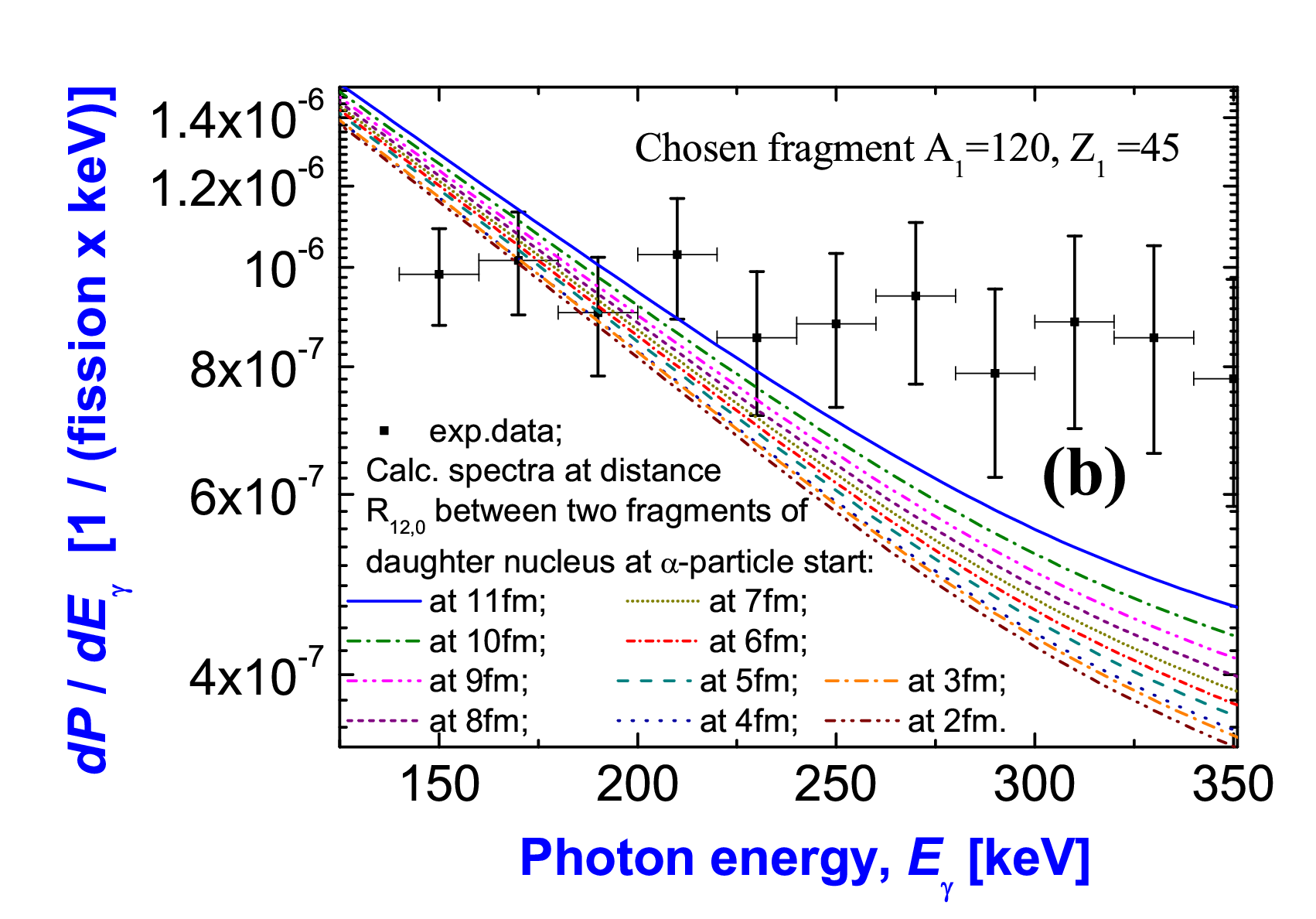}}
\vspace{-4mm}
\caption{\small (Color online)
The bremsstrahlung probabilities in  ternary fission of the \isotope[252]{Cf} nucleus when we use the mass separation of the binary nucleus into two fragments with $A_{1}=120$, $Z_{1}=45$ and $A_{2}=128$, $Z_{2}=51$; the other parameters used in these calculations are $\varphi=90^{\circ}$, and $C_{1}=0.3$.
The preliminary experimental data~\cite{Maydanyuk_2011} are also added on these figures.
(a) The full spectrum including the emission of the $\alpha$\,particle in the field of the binary nucleus and contribution of the emission of the two fragments of  binary nucleus by using the approximation (\ref{eq.5.2.4}) and for the function $F$ the form (\ref{form31});
(b) the spectra obtained for different distances $R_{12,\,0}$ between the two fragments of the binary nucleus at starting of emission of the $\alpha$\,particle. Full line for the $R_{12,\,0}$ distance of 11~fm, dotted line for  9~fm, dashed line for 7~fm, long dashed line for 5~fm, dash-dotted line for 3~fm, dash-double dotted line for 2~fm .
\label{fig.7}}
\end{figure}
Taking into account such a scheme, we assume for the function $F\,[k, \cos(\mathbf{k}, \mathbf{r})]$  the following form
\begin{equation}
F = 1 - f_{0}/(1 + \exp((w - w_{0})/w_{1})),
\label{form31}
\end{equation}
where $w$ is energy of photon. The unknown parameters $f_{0}$, $w_{0}$ and $w_{1}$ can be determined by comparing the theoretical calculations with  the experimental data in the whole energy region of photons. Applying the method above to the experimental data shown in Fig.~5 of Ref.~\cite{Maydanyuk_2011}, we find 
$f_{0} = 1.02$,
$w_{0} = 0.238$,
$w_{1} = 0.057$.

The calculated bremsstrahlung probability on the basis of such a scheme is shown in Fig.~\ref{fig.7}~(a) by the green line. One can see that the new spectrum, 
calculated on the basis of the improved approximation
(\ref{eq.5.2.4}) and the function $F\,[k, \cos(\mathbf{k}, \mathbf{r})]$, is 
highly sensitive on the parameters and the model describes mechanisms of the ternary fission process.
Thus, this dynamical model of the ternary fission describes experimental data with good accuracy.
Of course, to appropriately describe  the effects of the dynamics of the ternary fission regarding the bremsstrahlung photon emission, it is necessary to compare the calculation with reliable experimental data.

According to the logic above,
the result of Fig. \ref{fig.7}~(a) shows that 
at higher energies (above 300~keV) the emission of photons is mainly determined by contribution of two moving fragments of the binary nucleus
in the second stage of ternary fission process.
At lower energies, 
the emission of photons is caused by the $\alpha$\,particle in the field of the binary nucleus (before its separation of fragments) in the first stage of the ternary fission process. 
In other words, at energies smaller than 300~keV the analysis of the spectrum dependence on the distance $R_{12,\,0}$ between centers of masses of the  two fragments of the binary nucleus at starting point of emission of the $\alpha$\,particle is reliable. If we consider different possible values of the distances $R_{12,\,0}$ from 2 to 11 fm, we obtain for the photon emission probability the results which are reported in  Fig.~\ref{fig.7}~(b). As one can see, the present model is able to show by Figs. 4-7 the sensitivity of results  with respect to the values of the parameters describing  the ternary fission process mechanism.   From these results we can extract the optimal value of the $R_{12,\,0}$ distance when the calculated curve will be the closest to the available experimental data, at separation of the binary nucleus into two fragments with specific $A_{1}=120$, $Z_{1}=45$ and $A_{2}=128$, $Z_{2}=51$ values or by the prevailing mass distribution of  fragments.
Also note that possibly higher values of $R_{12,\,0}$ put it outside the region of applicability of the proposed approximations (\ref{eq.2.3.2.1}) and  (\ref{eq.2.3.3.1}) and convergence of current computer calculations of the spectra.



Finally, let us collect main aspects of the model with high influence on the bremsstrahlung probability. These are:
(i) different mass separations of 
the binary nucleus on 
two fragments;
(ii) the geometry of the fissioning nucleus;
(iii) the neck 
between two fragments of the binary nucleus at moment of separation of the $\alpha$\,particle;
(iv) the dynamics of relative motion of the $\alpha$\,particle and fragments of the binary nucleus;
(v) 
the different angles between the emitted photon and fission axis.
The motion of two fragments of the binary nucleus during fission gives an additional contribution to the photon emission. 
The last process turns out to have the highest influence on the bremsstrahlung.



\section{Conclusions and perspectives
\label{sec.conclusions}}

Bremsstrahlung in nuclear reactions provides useful information about 
mechanisms of reactions, structure of nuclei, nuclear interactions, etc..
Bremsstrahlung photons is subject of intensive investigations in nuclear physics for about 90 years.
As is known, the ternary fission is accompanied by emission of bremsstrahlung photons. 
However, such emission has never been studied theoretically and experimentally yet for the ternary fission.
So, in this paper we at first time clarify which new information about ternary fission can be obtained from 
study of
bremsstrahlung emission accompanying this process.
We developed 
a new quantum model of emission of bremsstrahlung photons accompanying ternary fission of heavy nuclei
with $\alpha$-particle as light charged particle. 
The model takes into account geometry and dynamics of the ternary fission.
%
By the theoretical results reported here, we establish high sensitivity of the bremsstrahlung spectra calculated by the model concerning to the following aspects of the ternary fission:

\begin{itemize}
\item[(a)]
Bremsstrahlung photons are emitted with the highest intensity in case of perpendicular motion of the $\alpha$\,particle concerning to fission axis
[see Fig.~\ref{fig.4}~(a)]. 
This direction of the emitted $\alpha$\,particle is reported in literature as the most probable case in the ternary fission~\cite{Hwang:2000nt}.

\item[(b)]
Relative motion between heavy fragments gives the highest contribution of bremsstrahlung to the full spectrum in the ternary fission.
Contribution of bremsstrahlung from process of leaving of $\alpha$-particle concerning to system of heavy fragments is significantly smaller. In addition, relative motion between two heavy fragments is faster, bremsstrahlung is more intensive [see Fig.~\ref{fig.4}~(b)]. 

\item[(c)]
The smallest intensity of bremsstrahlung in ternary fission corresponds to the case of a not separated binary nucleus on two heavy fragments. 

\item[(d)]
Small dependence of bremsstrahlung spectrum on geometry of neck between two heavy fragments 
(i.e., distance $R_{12,\,0}$ between the two fragments) at moment of separation of $\alpha$ particle is established
[see Fig.~\ref{fig.7}~(b)].


\end{itemize}


In summary, study of bremsstrahlung emission in ternary fission
shows high sensitivity of bremsstrahlung spectra on the geometry and dynamics of ternary fission process.
Good agreement between the calculated spectrum and preliminary experimental data~\cite{Maydanyuk_2011} 
is obtained 
on the basis of the new dynamical model 
[see Fig.~\ref{fig.7}~(a) and Sect.~ III.A].
On such a basis, new information can be obtained by our model after new experimental measurements of bremsstrahlung in this process.

\begin{acknowledgments}
We are highly appreciated to
Profs. Giorgio~Giardina and Giuseppe~Mandaglio for fruitful discussions concerning to aspects of nuclear reactions and bremsstrahlung,
Dr. Cheng Chen for interesting discussions concerning to nuclear reactions.
This work is partly supported by the National Key R\&D Program of China under Grant No. 2023YFA1606703, and by the National Natural Science Foundation of China under Grant Nos. 12435007 and 12361141819.

\end{acknowledgments}

\appendix
\section{Formalism on the contribution caused by the mass separation of the binary nucleus
\label{sec.app.1}}

Let us write the operator of the photon  emission of the system formed by the $\alpha$\,particle and  binary nucleus in the laboratory system, where the $\alpha$\,particle and  nucleus are composed of nucleons:
\begin{equation}
\begin{array}{lcl}
  \hat{H}_{\gamma} =
    - \displaystyle\sum\limits_{i=1}^{4}
    \displaystyle\frac{e\,Z_{i}}{m_{i}}\;
    \mathbf{A}(\mathbf{r}_{i},t)\, \mathbf{\hat{p}}_{i} -
    \displaystyle\sum\limits_{j=1}^{A}
    \displaystyle\frac{e\,Z_{j}}{m_{j}}\;
    \mathbf{A}(\mathbf{r}_{j},t)\, \mathbf{\hat{p}}_{j},
\end{array}
\label{eq.app.2.1}
\end{equation}
with
\begin{equation}
\begin{array}{lcl}
  \mathbf{A}(\mathbf{r},t) =
    \sqrt{\displaystyle\frac{2\pi}{w}} \;
    \displaystyle\sum\limits_{\eta=1,2} \mathbf{e}^{(\eta),*}
    e^{-i \mathbf{k\cdot r}}.
\end{array}
\label{eq.app.2.2}
\end{equation}

The coordinates of center-of-mass as $\mathbf{r}_{\alpha}$ for  $\alpha$\,particle, $\mathbf{R}_{A}$ for the binary nucleus and $\mathbf{R}$ for the complete system are given by
\begin{equation}
\begin{array}{lll}
  \mathbf{r}_{\alpha} = \displaystyle\frac{1}{m_{\alpha}} \displaystyle\sum_{i=1}^{4} m_{i}\, \mathbf{r}_{\alpha i}, \\

  \mathbf{R}_{A}      = \displaystyle\frac{1}{M} \displaystyle\sum_{j=1}^{A} m_{j}\, \mathbf{r}_{A j}, \\

  \mathbf{R}          = \displaystyle\frac{M\mathbf{R}_{A} + m_{\alpha}\mathbf{r}_{\alpha}}{M+m_{\alpha}}.
\end{array}
\label{eq.app.2.4}
\end{equation}
By introducing the relative coordinates $\mathbf{s}_{\alpha i}$, $\mathbf{s}_{A j}$ and $\mathbf{r}$, and using relations (\ref{eq.2.4.2}), (\ref{eq.2.4.3}) and  (\ref{eq.2.4.4} ) we find for the matrix element of photon emission the following form
\begin{widetext}
\begin{equation}
\begin{array}{llll}
  \vspace{2mm}
  \hat{H}_{\gamma} & = &
  -\,e\, \sqrt{\displaystyle\frac{2\pi}{w}}\,
    \displaystyle\sum\limits_{\eta=1,2} \mathbf{e}^{(\eta),*}\;
    e^{-i \mathbf{k} \cdot \Bigl[\mathbf{R} - \displaystyle\frac{m_{\alpha}}{M+m_{\alpha}}\,\mathbf{r} \Bigr]}\,

    \Biggl\{
      \Bigl[
        e^{-i \mathbf{k}\mathbf{r}}\,
          \displaystyle\sum\limits_{i=1}^{4}  \displaystyle\frac{Z_{i}}{m_{i}}\;
          e^{-i \mathbf{k} \cdot \mathbf{s}_{\alpha i} }\, +\,

          \displaystyle\sum\limits_{j=1}^{A}  \displaystyle\frac{Z_{j}}{m_{j}}\;
          e^{-i \mathbf{k}\cdot \mathbf{s}_{A j} }
      \Bigr]\: \mathbf{P} + \\

  & & 
    \Bigl[
    e^{-i \mathbf{k}\mathbf{r}}\,
          \displaystyle\frac{M}{M+m_{\alpha}}
          \displaystyle\sum\limits_{i=1}^{4} \displaystyle\frac{Z_{i}}{m_{i}}\;
          e^{-i \mathbf{k}\cdot \mathbf{s}_{\alpha i} }\, -
        \displaystyle\frac{m_{\alpha}}{M+m_{\alpha}}\:
          \displaystyle\sum\limits_{j=1}^{A} \displaystyle\frac{Z_{j}}{m_{j}}\;
          e^{-i \mathbf{k}\cdot \mathbf{s}_{A j} }
      \Bigr]\: \mathbf{p} + 
      e^{-i \mathbf{k}\mathbf{r}}\,
      \displaystyle\sum\limits_{i=1}^{4}
        \displaystyle\frac{Z_{i}}{m_{i}}\;
        e^{-i \mathbf{k}\cdot \mathbf{s}_{\alpha i} }\, \mathbf{p}_{\alpha i}\: +\:
      \displaystyle\sum\limits_{j=1}^{A}
        \displaystyle\frac{Z_{j}}{m_{j}}\;
        e^{-i \mathbf{k}\cdot \mathbf{s}_{A j}}\, \mathbf{p}_{A j}
    \Biggr\},
\end{array}
\label{eq.app.2.9}
\end{equation}
\end{widetext}
where we used the following operators:
\begin{equation}
\begin{array}{llll}
  \vspace{2mm}
  \mathbf{p} = -i \displaystyle\frac{\mathbf{d}}{\mathbf{dr}}, &
  \mathbf{P} = -i \displaystyle\frac{\mathbf{d}}{\mathbf{dR}}, \\
  \mathbf{p}_{\alpha i} = -i \displaystyle\frac{\mathbf{d}}{\mathbf{dr}_{\alpha i}}, &
  \mathbf{p}_{Aj} = -i \displaystyle\frac{\mathbf{d}}{\mathbf{dr}_{Aj}}.
\end{array}
\label{eq.app.2.9b}
\end{equation}

Following Ref. \cite{Kopitin.1997.YF} we present the wave function of total system by the form:
\begin{equation}
  | \phi\rangle = e^{-i\,\mathbf{K}_{\phi}\cdot\mathbf{R}}\: \phi_{\alpha}\; |\,\phi_{A} \rangle,
\label{eq.app.2.10}
\end{equation}
where $\phi = i$ or $f$ (indices $i$ and $f$ denote the initial state (before photon emission) and the final state (after photon emission)), $\mathbf{K}_{\phi}$ is the full momentum of the total system, $|\,\phi_{A}\rangle$ is the  wave function describing the internal states of the binary nucleus, $\phi_{\alpha}$ is the wave function describing the relative motion (with possible tunneling) of $\alpha$\,particle concerning the binary nucleus.
Suggesting
%
$
  \mathbf{K}_{i} = 0,
$
%
we calculate the matrix element:

\begin{widetext}
\begin{equation}
\begin{array}{lcl}
  \vspace{0mm}
  \langle f |\, \hat{H}_{\gamma} |\,i \rangle & = &
  -\,e\, \sqrt{\displaystyle\frac{2\pi}{w}}\,
    \displaystyle\sum\limits_{\eta=1,2} \mathbf{e}^{(\eta),*}\;
  \Biggl\{
  \Biggl\langle
    f_{\alpha},\, f_{A}\,
  \Biggl|\,
    e^{i\,(\mathbf{K}_{f} - \mathbf{k})\cdot\mathbf{R}}\:
    e^{i\, \mathbf{k\cdot r}\, \displaystyle\frac{m_{\alpha}}{M+m_{\alpha}}}\, \times \\

  \vspace{2mm}
  & & \times\;
      \Bigl[
        e^{-i \mathbf{k}\cdot \mathbf{r}}\,
          \displaystyle\sum\limits_{i=1}^{4}  \displaystyle\frac{Z_{i}}{m_{i}}\;
          e^{-i \mathbf{k}\cdot \mathbf{s}_{\alpha i} }\, +
          \displaystyle\sum\limits_{j=1}^{A}  \displaystyle\frac{Z_{j}}{m_{j}}\;
          e^{-i \mathbf{k}\cdot \mathbf{s}_{A j} }
      \Bigr]\: \mathbf{P}\;
    \Biggr|\,
      i_{\alpha},\, i_{A}\,
    \Biggr\rangle + \\

  & & +\:
    \Biggl\langle
      f_{\alpha},\, f_{A}\,
    \Biggl|\,
      e^{i\,(\mathbf{K}_{f} - \mathbf{k})\cdot\mathbf{R}}\:
      e^{i\, \mathbf{k\cdot r}\, \displaystyle\frac{m_{\alpha}}{M+m_{\alpha}}}\,
      \Bigl[
        e^{-i \mathbf{k}\cdot \mathbf{r}}\,
          \displaystyle\frac{M}{M+m_{\alpha}}
          \displaystyle\sum\limits_{i=1}^{4} \displaystyle\frac{Z_{i}}{m_{i}}\;
          e^{-i \mathbf{k}\cdot \mathbf{s}_{\alpha i} }\, - \\
        & & \qquad\qquad\qquad\qquad -\;
        \displaystyle\frac{m_{\alpha}}{M+m_{\alpha}}\:
          \displaystyle\sum\limits_{j=1}^{A} \displaystyle\frac{Z_{j}}{m_{j}}\;
          e^{-i \mathbf{k}\cdot \mathbf{s}_{A j} }
      \Bigr]\: \mathbf{p}
    \Biggr|\,
      i_{\alpha},\, i_{A}\,
    \Biggr\rangle\; + \\

  \vspace{2mm}
  & & +\:
    \Biggl\langle
      f_{\alpha},\, f_{A}\,
    \Biggl|\,
      e^{i\,(\mathbf{K}_{f} - \mathbf{k})\cdot\mathbf{R}}\:
      e^{i\, \mathbf{k\cdot r}\, \displaystyle\frac{m_{\alpha}}{M+m_{\alpha}}}\,
      e^{-i \mathbf{k}\cdot \mathbf{r}}\,
      \displaystyle\sum\limits_{i=1}^{4}
        \displaystyle\frac{Z_{i}}{m_{i}}\;
        e^{-i \mathbf{k}\cdot \mathbf{s}_{\alpha i} }\, \mathbf{p}_{\alpha i}
    \Biggr|\,
      i_{\alpha},\, i_{A}\,
    \Biggr\rangle\; + \\
  & & +\;
    \Biggl\langle
      f_{\alpha},\, f_{A}\,
    \Biggl|\,
      e^{i\,(\mathbf{K}_{f} - \mathbf{k})\cdot\mathbf{R}}\:
      e^{i\, \mathbf{k\cdot r}\, \displaystyle\frac{m_{\alpha}}{M+m_{\alpha}}}\,
      \displaystyle\sum\limits_{j=1}^{A}
        \displaystyle\frac{Z_{j}}{m_{j}}\;
        e^{-i \mathbf{k}\cdot \mathbf{s}_{A j}}\, \mathbf{p}_{A j}\;
    \Biggr|\,
      i_{\alpha},\, i_{A}\,
    \Biggr\rangle
    \Biggr\}.
\end{array}
\label{eq.app.2.12}
\end{equation}
\end{widetext}

The effective charge of the nuclear system ($\alpha$\,particle and binary nucleus) is written as
\begin{equation}
\begin{array}{lcl}
\vspace{1mm}
  Z_{\rm eff}(\mathbf{r}) & = &
    e^{i\,\mathbf{k\cdot r}}\:
    \biggl\{
      e^{- i\,\mathbf{k\cdot r}\, \displaystyle\frac{M}{M+m_{\alpha}}}\:
        \displaystyle\frac{M\, Z_{\rm \alpha}}{M+m_{\alpha}} - \\
  & &
  e^{i\,\mathbf{k\cdot r}\, \displaystyle\frac{m_{\alpha}}{M+m_{\alpha}}}\:
 \displaystyle\frac{m_{\alpha}\, Z_{\rm A}}{M+m_{\alpha}}
    \biggr\}.
\end{array}
\label{eq.app.2.13}
\end{equation}
The charged form-factor of $\alpha$\,particle is given by
\begin{equation}
  Z_{\rm \alpha} (\mathbf{k}) =
  \Bigl\langle \beta_{f} \Bigl|
    \displaystyle\sum\limits_{i=1}^{4} \displaystyle\frac{Z_{i}}{m_{i}}\;
    e^{-i \mathbf{k}\cdot \mathbf{s}_{\alpha i} } , 
  \Bigr|\, \beta_{i} \Bigr\rangle
\label{eq.app.2.14}
\end{equation}
and the charged form-factor of the  binary nucleus is
\begin{equation}
  Z_{\rm A} (\mathbf{k}) =
  \Bigl\langle \beta_{f} \Bigl|\,
    \displaystyle\sum\limits_{j=1}^{A} \displaystyle\frac{Z_{j}}{m_{j}}\; e^{-i \mathbf{k}\cdot \mathbf{s}_{A j} }
  \Bigr|\, \beta_{i} \Bigr\rangle.
\label{eq.app.2.15}
\end{equation}
Then, we obtain:
%
\begin{equation}
\begin{array}{lll}
\vspace{1.5mm}
  \langle f |\, \hat{H}_{\gamma} |\,i \rangle =
  -\,e\, \sqrt{\displaystyle\frac{2\pi}{w}}\,
    \displaystyle\sum\limits_{\eta=1,2} \mathbf{e}^{(\eta),*}\;
    \delta(\mathbf{K}_{f} - \mathbf{k}) \times \\

\vspace{1.5mm}
  \biggl\{
      \biggl\langle f_{A},\, f_{\alpha} \biggl|\,
        Z_{\rm eff}(\mathbf{r})\: e^{-i\,\mathbf{k\cdot r}}\: \mathbf{p}\;
      \biggr|\,
        i_{A},\, i_{\alpha}
      \biggr\rangle\; + \\
    \Biggl\langle
      f_{\alpha}
      \Biggl|\,
        e^{i\, \mathbf{k\cdot r}\, \displaystyle\frac{m_{\alpha}}{M+m_{\alpha}}}\,
        \Bigl\langle\, f_{A} \Bigl|\,
          Z_{\rm A}(\mathbf{k})\, \mathbf{p}_{A j}\;
        \Bigr|\, i_{A} \Bigr\rangle\,
      \Biggr|\,
      i_{\alpha}
    \Biggr\rangle
    \Biggr\}.
\end{array}
\label{eq.app.2.16}
\end{equation}
%
Note that in the first approximation (called dipole) one can get,
\begin{equation}
  Z_{\rm eff} =
  \displaystyle\frac{M\, Z_{\rm \alpha} - m_{\alpha}\, Z_{\rm A}}{M+m_{\alpha}}.
\label{eq.app.2.17}
\end{equation}



\end{document}